\documentclass{aa}  

\usepackage{graphicx}
\usepackage{wasysym}
\usepackage{amsmath}
\usepackage{breqn}
\usepackage{txfonts}
\usepackage{subfigure}
\usepackage{color}
\usepackage{tikz}
\usetikzlibrary{positioning}
\usepackage{longtable} 
\usepackage{multirow}
\usepackage{pdflscape}
\usepackage[hyphenbreaks]{breakurl}
\usepackage[breaklinks]{hyperref}
\hypersetup{pdfauthor=Remo Burn, pdftitle=Formation of Water Worlds, breaklinks=true, colorlinks=true, urlcolor=blue, linkcolor=blue, citecolor=blue, bookmarksopen=true}

\usepackage{natbib}

\usepackage[separate-uncertainty]{siunitx}
\DeclareSIUnit{\mearth}{M_\oplus}
\DeclareSIUnit{\mjup}{M_{Jup}}
\DeclareSIUnit{\rearth}{R_\oplus}
\DeclareSIUnit{\rjup}{R_\jupiter}
\DeclareSIUnit{\rsol}{R_{\odot}}
\DeclareSIUnit{\msol}{M_{\odot}}
\DeclareSIUnit{\au}{au}
\DeclareSIUnit{\year}{yr}
\DeclareSIUnit\erg{erg}

\renewcommand*\d[0]{\mathrm{d}}

\begin{document}

   \title{Water-rich sub-Neptunes and rocky super Earths around different Stars: Radii shaped by Volatile Partitioning, Formation, and Evolution}
   \titlerunning{Water-rich sub-Neptunes and rocky super Earths around different stars}
   \authorrunning{Burn et al.}

   \author{Remo Burn
          \inst{1,2} \and
          Komal Bali \inst{3} \and
          Caroline Dorn \inst{3} \and
          Rafael Luque \inst{4,5,6} \and
          Simon L. Grimm \inst{7,3}
          }

   \institute{Max Planck Institute for Astronomy, K\"onigstuhl 17, 69117 Heidelberg, Germany
   	      \and
          Observatoire de la C\^ote d'Azur, 96 Bd de l'Observatoire, 06300 Nice, France\\
          \email{remo.burn@oca.eu}
          \and Institute for Particle Physics and Astrophysics, ETH Z\"urich, HPK E32, Otto-Stern-Weg 5, 8093 Z\"urich,Switzerland
         \and  Department of Astronomy \& Astrophysics, University of Chicago, Chicago, IL 60637, USA
         \and NHFP Sagan Fellow
         \and Instituto de Astrof\'isica de Andaluc\'ia (IAA-CSIC), Glorieta de la Astronom\'ia s/n, 18008 Granada, Spain
         \and University of Z\"urich, Department of Astrophysics, Winterthurerstrasse 190, 8057 Z\"urich, Switzerland
             }

   \date{\today}

  \abstract
   {
   Despite precise characterization measurements, the nature of planets with radii ranging from 2 to 4 Earth radii, the sub-Neptunes, remains unknown due to degeneracies in interior models. However, an impressive ensemble of small planets with measured masses and radii around different stars has been compiled by the field.
   }
   {It can be used to test the prediction of large water reservoirs on sub-Neptunes by planet formation theory with orbital migration. We want to find out whether this water reservoir is included in photoevaporative winds and how much of it can partition into the rocky and metallic interior.}
   {We couple the result of a planetary formation model with planetesimal and gas accretion as well as orbital migration to evolution models which assume perfect mixing of water with H/He in the envelope or complete segregation. For the mixed envelopes, we also include an analytic treatment of fractionation during photoevaporative mass-loss. Further, the effect of equilibrium dissolution of water into an assumed magma ocean and into the metallic core is studied for the first time in coupled formation-evolution models.}
   {Out of the tested scenarios, the mass-radius relation of exoplanets is relatively well matched by all scenarios where the atmosphere is of mixed composition. The agreement depends on mass, with better consistency for the model without dissolution below 3 Earth masses and hints of the opposite at higher masses. Fractionation is not found to significantly alter the properties of the planets for our initial conditions due to initially massive envelopes on all planets. For all scenarios, we quantify the radius valley location and scaling with stellar mass.}
   {Water-rich sub-Neptunes mass-radius relations are broadly consistent with observations but statistical surveys in mass and radius are required for distinction of the scenarios. The mixed-composition loss of different volatiles, their dissolution into the planetary interior, and the solidification of the magma ocean are natural next steps toward a comprehensive atmosphere-interior evolution model.}

   \keywords{Planets and satellites: formation -- Planets and satellites: physical evolution -- Planets and satellites: interiors -- Planets and satellites: atmospheres -- Planet-star interactions}

   \maketitle

\section{Introduction}
So far, the mass, orbital period, and radius of around two hundred sub-Saturn-mass exoplanets have been characterized with high precision \citep{Parc2024}. This success was possible thanks to various observational programs, such as K2 \citep{Howell2014} or TESS \citep{Ricker2015}, which find nearby planets amenable to characterization via photometric and radial-velocity follow-up observations \citep{Astudillo-Defru2020,Bonomo2023,Bonfanti2024,Brinkman2023,Damasso2023a,Cloutier2024,Giacalone2022,Kossakowski2021,Leleu2024,Luque2022,Luque2022b,Luque2023,Orell-Miquel2023,Osborn2021,Passegger2024,Ulmer-Moll2023}. The set of exoplanet data complements the statistical study of exoplanet occurrence from Kepler \citep{Borucki2010,Lissauer2023} where the masses of the planets are often unconstrained.

But even with a precisely determined mass, the interior structure of the planet is not fully characterized. This is the case for rocky planets due to an a priori unknown metallic core to rocky mantle ratio \citep{Dorn2015} and even more so if volatile species such as water, hydrogen (H), and helium (He) are present. Without additional constraints from atmospheric spectroscopy, this degeneracy is impossible to break for an individual object. However, the statistical study of the exoplanet demographics can reveal further insights into the typical structure of an average exoplanet. One key feature which was discovered is the so-called radius valley which separates the smaller super Earths from the larger sub-Neptunes \citep{Fulton2017,Fulton2018,VanEylen2018}. For planets with gaseous envelopes which are continuously stripped by photoevaporation, driven by high-energy radiation from the star, such a transition was predicted to exist \citep{Owen2013,Lopez2013,Jin2014,Chen2016} and offered a compelling explanation. With the increasing knowledge of the mass and thus the bulk density of these planets, \citet{Luque2022} found that -- at least for M-dwarf planets -- the radius valley as seen by Kepler is fundamentally the projection of a gap in planetary bulk density, as expected from a compositional transition.

An open question remains: what is the volatile material present on sub-Neptunes but absent on super Earths? Traditionally, it was thought to be pure H/He and large quantities of water ice were excluded \citep{Lopez2017,Jin2018,Rogers2021a}. However, water accreted as ice likely changes its phase after accretion and could be present in large quantities as steam (partially) mixed with H/He \citep{Dorn2018,Jin2018,Zeng2019,Mousis2020,Venturini2020,Turbet2020,Acuna2021,Aguichine2021,Pierrehumbert2023}. Here, we will explore scenarios where water mixes with H/He, as well as a case with two distinct, stratified layers.

If water accounts for a significant portion (on the order of tens of percent) of the total mass of the planet, it likely accreted it in solid form beyond the water iceline in the protoplanetary disk. Therefore, the presence or absence of water on sub-Neptunes is a fundamental probe for large scale migration of these planets \citep{Lin1996,Ward1997,Paardekooper2023} and provides essential constraints for modern planet formation modeling. The orbits of sub-Neptune pairs are commonly in mean-motion resonance \citep{Leleu2024a,Chen2024}, which is likely linked to migration. However, it can be reproduced by small-scale convergent migration. Verifying the migration hypothesis through compositional analysis would prove that migration of sub-Neptunes over significant distances occurred.

Indeed, the limited masses reached by purely rocky planets forming via core accretion strongly favors that the origin of sub-Neptune mass planets lies at larger distances \citep{Alibert2005,Emsenhuber2023a}. Therefore, a water-rich composition is the focus of studies coupling formation and evolution models to explore the radius valley \citep{Venturini2020,Chakrabarty2024,Burn2024}.

Additionally, early evidence from JWST transmission spectroscopy suggests that some of the observed sub-Neptune atmospheres are likely water-rich. So far, the presence of water was confirmed in TOI-270~d \citep{Holmberg2024,Benneke2024}, GJ 3470 b \citep{Beatty2024}, and GJ 9827~d \citep{Piaulet-Ghorayeb2024}, the latter begin consistent with a water-rich ''steam world''. Tentative detections have been obtained for other sub-Neptunes as well \citep{MacDougall2023,Kempton2023,Wallack2024}. Further observations are scheduled for this population, which have the potential to quantify the atmospheric water abundance \citep{Kempton2024,Nixon2024}. Combined with data on the bulk properties, these measurements help to reveal the internal composition of sub-Neptunes.

If the atmosphere consists of more than one species, it is possible that atmospheric escape preferentially leads to loss of the lighter species relative to the heavier species. If hydrogen is lost in larger abundance than the heavier species, this would change the metallicity in the envelope over time and could lead to a metal-rich late atmosphere. This process, called fractionation, was first studied analytically \citep{Hunten1987,Zahnle1986,Zahnle1990} and is a common subject in the context of Solar System planets' atmospheres \citep{Chassefiere1996,Odert2018,Lammer2020}. For exoplanets, the process was investigated for the fractionation of He from H \citep{Hu2015,Malsky2020,Cherubim2024} and found large He fractions in some cases. \citet{Gu2023} reported significant deuterium fractionation on sub-Neptunes occurring in a part of the parameter space. So far, no studies address the fractionation of heavier elements in sub-Neptune atmospheres. In this work, we investigate using a simple analytic approach the fractionation of oxygen (from assumed dissociated, abundant water molecules) from H, which is a required addition for water-rich sub-Neptunes.

In addition to the exact atmospheric partitioning, there is another large source of uncertainty on the water partitioning in the planet \citep[see][for recent reviews]{Ikoma2018,Lichtenberg2023}. Instead of mixing with the atmosphere, water can be sequestered in the deep interior (mantle and core). While solidified rocky mantle material can only hold limited amounts of volatiles \citep{Shah2021}, large quantities of volatiles can be dissolved in magma \citep{Hirschmann2012,Dorn2021,Sossi2023}. Moreover, during core formation, volatiles can also get sequestered in the iron core \citep{Luo2024}. At temperatures above 2000\,K, supercritical rock and water can mix \citep{Vazan2022}. The presence of volatiles in the deeper interior decreases their amount in the envelope. As a consequence, the total radius and composition change, with interesting effects on the evolution of the planet, especially on photoevaporative mass loss, which is sensitive to both properties. Evolution models coupled to the interior were developed for the case of (mostly) hydrogen dissolution into the magma ocean by \cite{Chachan2018,Kite2019,Kite2020,Lichtenberg2021b,Schlichting2022} who found significant impacts on the resulting radii. In particular, \citet{Kite2019} found a truncation of the radius distribution at the upper end of the sub-Neptune radii (the radius cliff) when considering the non-ideality of the hydrogen in contact with the magma which leads to more mass being dissolved. Recently, \citet{Rogers2024a} developed an evolution model for super Earths including the effect of water dissolution into magma \citep[see also][]{Kite2021} and iron core following \citet{Luo2024}. They use it to infer allowed water mass fractions for super Earths and generally treat the initial conditions as free parameters. Here, we present a complementary model but use initial conditions from formation modeling instead to predict the effect of water sequestration on the population of simulated planets.

Another aspect of exoplanet demographics is the dependency on stellar properties, in particular the stellar mass \citep[but see also][for the dependency on metallicity]{Chen2025}. Several studies have focused on stellar mass dependency of the radius valley \citep{Berger2020a,Cloutier2020,VanEylen2021,Petigura2022,Luque2022,Ho2024,Bonfanti2024,Parc2024} and agree in inferring the locus of the valley at lower radii around smaller stars as predicted by theoretical works \citep{Gupta2020,Rogers2023}. This trend was tested including water-rich atmospheres in \citet{Venturini2024} using the prediction of a formation model where pebble accretion is the mode of solid mass delivery. They confirm a persistent density valley and report a contamination of the radius valley due to lower-mass water worlds around low-mass stars. Previously, scatter in these relations was often linked to collisions \citep{Inamdar2015,Izidoro2022}. \citet{Venturini2024} include the effect of collisions but rely on estimates using single-planet formation modeling. Here, we use results from full N-body interactions between 50 growing planets per disk. However, we use a simple treatment of collision outcomes, slightly modified compared to our preceding study focusing on Solar-mass stars \citep{Burn2024}.

In this paper, we include for the first time in coupled formation and evolution models 1) the effects of fractionation between oxygen and hydrogen during mass loss, and 2) water sequestration in the interior. We proceed by summarizing previously used methods and introducing in detail the new aspects in Sect. \ref{sec:methods} before presenting our results in different dimensions in Sect \ref{sec:results}. The results motivate some discussion (Sect. \ref{sec:discussion}) on the best way to compare distributions in mass-radius space with observations (Sect. \ref{sec:fits}), the unexpectedly weak impact of fractionation (Sect. \ref{sec:fractionation_interpretation}), and a possible dissimilarity between observational data and the model results including water sequestration (Sect. \ref{sec:sequestration_discussion}). Finally, we will summarize and conclude our findings in Sect. \ref{sec:conclusion}.

\section{Methods}
\label{sec:methods}
\subsection{Formation}
\label{sec:formation}
The (exo-)planetary mass and radius distribution are shaped by both the embedded stage in the protoplanetary disk as well as the long-term evolution of each planet. For the former, which we call here the formation stage, we use the results of the population synthesis work by \citet{Burn2021}. In that work, for five different stellar masses (0.1, 0.3, 0.5, 0.7, and 1.0\,M$_\odot$), an initial protoplanetary disk with 50 embedded, moon-mass embryos was assumed. Since the protoplanetary disk properties are expected to depend on the stellar mass, the initial disk masses following the observations of Class I disks by \citet{Tychoniec2018} were scaled linearly with the stellar mass. Similarly, the disk radius decreases for lighter disks \citep[$\propto M_{\rm gas}^{0.625}$]{Andrews2010,Andrews2018} while the disk lifetime, solid mass fraction, and inner edge orbital period were chosen not to depend on stellar mass. In \citet{Emsenhuber2023}, it was confirmed to a large degree that such initial conditions can lead to the observed, evolved Class II disk properties. The approach to randomize those initial conditions allows for statistical comparison of the resulting planets with observations, which is the idea of population syntheses \citep{Mordasini2024,Burn2024a}. The disk subsequently evolves, which is modeled in \citet{Burn2021} assuming a turbulent viscosity parameter $\alpha$ of \SI{2e-3}{} and a simple photoevaporation prescription tuned such that the disk lifetimes are distributed around 3\,Myr for all modeled stars \footnote{While there is a trend of shorter disk lifetimes around stars more massive than 1.5\,M$_\odot$ compared to those lighter than this threshold mass, this trend is not confirmed to exist below 1\,M$_\odot$ \citep{Richert2018}. See also the discussion in \citet{Burn2021}, but note the exception of some long-lived disks \citep{Silverberg2020}.}

The formation of planets as modeled by \citet{Burn2021} uses the model described in detail in \citet{Emsenhuber2020a} which is based on the works of \citet{Alibert2005,Mordasini2009,Mordasini2012c}. It includes the N-body forces between the embryos \citep[using \texttt{mercury,}][]{Chambers1999} and an analytically described accretion of planetesimals by a protoplanet with a gaseous envelope \citep{Inaba2001,Inaba2003,Fortier2013}, which dominate the first modeled stage of growth. At intermediate masses, the accretion of gas becomes important which is calculated by solving the one-dimensional internal structure equations \citep{Bodenheimer1986} of the gaseous envelope and tracking its energy content which allows the description of its cooling and contraction. As long as the planet's growth is not limited by the mass of the surrounding gaseous disk, the amount of gas accreted is given by the contraction of the envelope. When the planet mass becomes dominated by gas instead of solids, the larger compressibility of gas allows for increased contraction which leads to even more gas accretion. Thus, the planet enters a runaway accretion stage \citep{Makino1998}. This rapid gas accretion phase is stopped once the disk is not able to supply by viscous transport the amount of gas that could be accreted by the planet given its cooling and contraction rate. At this point, the planet detaches from the gaseous disk, contracts, and its accretion rate is modeled using an analytical estimate for the maximum gas accretion rate \citep{Emsenhuber2020a}.

While the disk is present, it will be perturbed by the young planet and exert a torque on it. Therefore, the planets orbital elements are changed and in our model follows the prescriptions of \citet{Paardekooper2011} for type I migration and \citet{Dittkrist2014} for type II migration. Migration will have the effect of typically moving volatile-rich planets formed beyond the snow line toward the star. Therefore, a common outcome to models including planetary migration is the prediction of planets with volatile (mostly water) contents of several tens of percent in bulk mass at short orbital periods \citep{Ida2008,Alibert2005}. This effect is even more pronounced for lower stellar masses due to the lack of planets reaching the slower type II regime and shorter distance from the snowline to the inner system \citep{Alibert2017,Burn2021}. Nevertheless, in the simulations of \citet{Burn2021}, which form the basis of our calculations, rocky planets emerging interior to the water iceline can also form due to an assumed high efficiency of planetesimal formation in this region. This is in contrast to other simulations assuming reduced efficiency of accretion of dry particles or different initial planetesimal distributions \citep{Coleman2019,Miguel2020}.

\subsection{Evolution}
After the planets have accreted solids and gas and might have migrated, the protoplanetary disk eventually dissipates. Starting from this point in time, defined by the local disk pressure around the planet dropping to low levels, we model the evolution of the planet as individual body. In this work, we modified the approach of \citet{Emsenhuber2020a} in a few key aspects similar to \citet{Burn2024} in order to better model planets in the sub-Neptune regime with large volatile contents. The main aspect is the used equation of state (EOS) for volatiles for which we use an EOS for water \citep{Haldemann2020} which covers all phases. For simplicity, we treat for the evolution phase all volatile species accreted as ice (i.e. H$_2$O, CO, CO$_2$, CH$_3$OH, CH$_4$, NH$_3$, N$_2$, and H$_2$S, see \citealp{Marboeuf2014b,Marboeuf2014}) as water ice and will use "water" and "volatiles" as synonyms in this work. 

We focus on four different evolution models which differ in their assumed underlying partitioning of water into the interior structure. The approaches are summarized in Figure \ref{fig:model_schematics} and are using: (1) the Mixed assumption, that is, the nominal model, where all volatiles treated as water are perfectly mixing with H and He, (2) a similar model but where the mixing assumption is relaxed for the lost mass due to photoevaporation, that is, Fractionation is allowed to occur, (3) the Layered model which does not mix the volatiles accreted as ice with other constituents, and finally (4) a model exploring the effect of dissolution of water to the magma ocean and metallic core, that is, the Water Sequestration model.

\subsubsection{Planetary structure and composition}
\label{sec:structure_and_compo}
\paragraph{General}
In all cases, we start at the disk dissipation time $t_{\rm disk}$ with initial conditions from the formation part of the models (Sect. \ref{sec:formation}). To improve the consistency, we recover the luminosities of the planets at $t_{\rm disk}$ analytically and rerun the evolution part from this time onward. A difference to \citet{Burn2024} is that we also use the hydrogen and helium (H/He) mass $M_{\rm H/He}$ from $t_{\rm disk}$ as starting condition instead of the one after \SI{20}{Myr} of evolution. This has an effect on simulated planets for which an impact occurred after $t_{\rm disk}$. By starting with $M_{\rm H/He}(t_{\rm disk}$, we assume that no H/He is lost for the more massive planet in a collision while all H/He is lost for the impacting planet. The mass of other components, that is iron, rocky materials, and volatile ices, is however assumed to perfectly merge. Future works will re-introduce a more realistic impact stripping with the possibility of mass loss of all species. Therefore, in the scenarios where we assume that volatile species mix with H/He (Mixed, Fractionation, see below) we obtain a total heavy element content (by mass) in the envelope of \begin{equation}
Z_{\rm homo} = \frac{M_{\rm vol}}{M_{\rm vol} + M_{\rm H/He}}
\end{equation}
smaller or equal to that in \citet{Burn2024}. 

Further common elements in all models are the presence of an iron core with a silicate MgSiO$_3$ (perovskite) mantle both materials described with a modified polytropic equation of state $\rho(P) = \rho_0 + c P^n$ of \citet{Seager2007}.

In all model cases, the uppermost -- third or fourth -- layer is what we call the envelope and is numerically resolved in one dimension. Similar to the formation stage \citep[described in][]{Emsenhuber2020a}, the following internal structure equations \citep[e.g.][]{Kippenhahn2012} in hydrostatic equilibrium are solved:
\begin{align}
\frac{\d m(r)}{\d r} &= 4 \pi r^2 \rho(r) \label{eq:dmdr}\\
\frac{\d P(r)}{\d r} &= - \frac{G m(r) \rho(r)}{r^2}\label{eq:dPdr}\\
\frac{\d T(r)}{\d r} &= \frac{T}{P} \frac{\d P}{\d r} \mathrm{min}(\nabla_{\mathrm{ad}},\nabla_{\mathrm{rad}})\label{eq:dTdr}\,,
\end{align}
where $m(r)$ is the mass within a sphere of radius $r$, $P(r)$ is the pressure as a function of $r$, $\rho(r)$ is the local density, and $T(r)$ the temperature. Depending on the local conditions, the energy transport mechanism changes. We use the limiting case of convection where the adiabatic gradient, calculated from the EOS,
\begin{equation}
\nabla_{\mathrm{ad}} = \left(\frac{\partial \ln T}{\partial \ln P}\right)_{S}
\label{eq:adgrad}
\end{equation}
is smaller than the radiative gradient. The latter depends on the local optical depth $\tau$. If $\tau \lesssim 10$, we use the atmosphere solution of \citet{Guillot2010}, which considers an outgoing infrared flux and a well separated incoming flux in the visible wavelength regime. This solution assumes both parts of the spectrum to be individually gray and further uses the Eddington approximation \citep[see also][]{Jin2014}. An important parameter in that model is the ratio of the opacities for the incoming (visible) and outgoing radiation (thermal). Here, we use linear interpolation within the values tabulated as a function of equilibrium temperature by \citet{Jin2014} in their Table 2. If the equilibrium temperature is outside the tabulated interval (260 to 2777\,K), we extrapolate linearly below 260\,K and quadratically in $T_{\rm eq}$ above 2777\,K. We emphasize that the opacity ratios were obtained from detailed radiation transfer models assuming Solar composition envelopes. If sub-Neptunes are water rich, we would expect stronger greenhouse effect due to high infrared opacities of water. In this work, we neglect this effect, which allows us to isolate the effects of photoevaporation, which are expected to dominate for the short-orbit planets considered here.

For the explicit dependency on optical depth in the solution of \citet{Guillot2010}, as well as for the radiative diffusion below, we use metallicity dependent opacities tabulated by \citet{Freedman2014} but note that there is a maximum tabulated metallicity of [M/H] =1.7, that is, it is not strictly applicable to $Z_{\rm env} \gtrsim 0.7$. Both the opacity and the incoming-to-outgoing opacity ratio should be generalized to differing compositions in future studies. In case of large optical depths, radiative diffusion is a well justified approximation, thus 
\begin{equation}
\nabla_{\mathrm{rad}} = - \frac{3}{16 \pi a c} \frac{\kappa L \rho}{r^2 T^3}
\end{equation} 
where $c$ is the speed of light in vacuum, $a = 4\sigma/c$ with the Stefan-Boltzmann constant $\sigma$. In an intermediate regime $14.4 < \tau < 144$, we linearly interpolate between the two regimes to find $\nabla_{\mathrm{rad}}$.

Following \citet{Mordasini2012c}, we assume that the luminosity $L$ is constant throughout the structure which reduces the number of equations and is justified as only the upper, radiative regime depends directly on the luminosity. The system of equations is closed by the EOS (see below). A shooting method is used to solve the boundary value problem which arises due to core size, mass, and temperature being given as lower boundary while external pressure and density are upper boundaries to the envelope structure. A dependency on the stellar type enters directly into the upper boundary temperature of the planetary structures in the form of the irradiation temperature $T_{\rm irr} = T_\star \sqrt{ R_\star / a_{\rm P}}$, where the stellar temperature $T_\star$ and radius $R_\star$ are taken from the tracks of \citet{Baraffe2015}.

Assuming that a hydrostatic equilibrium is established on short timescales, the structure of the planet is recalculated after each step forward in time. From the obtained pressure and temperature profile, the corresponding change in energy content (gravitational and internal) can then be estimated. This estimate can be used as the luminosity for the next timestep and leads to a method which was found to result in a consistent treatment of a cooling and contracting planet \citep{Mordasini2012c}. In addition to the energy budget from solid and gas accretion, we consider radiogenic heating from the decay of $^{40}$K, $^{232}$Th, and $^{238}$U \citep{Mordasini2012b} as an additional energy source. We further add an empirically determined source of energy causing the observed bloating of close-in planets following \citet{Sarkis2021}.

\begin{figure}
	\centering
	\def\svgwidth{.9\linewidth}
	\begingroup
	\makeatletter
	\providecommand\color[2][]{
		\errmessage{(Inkscape) Color is used for the text in Inkscape, but the package 'color.sty' is not loaded}
		\renewcommand\color[2][]{}
	}
	\providecommand\transparent[1]{
		\errmessage{(Inkscape) Transparency is used (non-zero) for the text in Inkscape, but the package 'transparent.sty' is not loaded}
		\renewcommand\transparent[1]{}
	}
	\providecommand\rotatebox[2]{#2}
	\newcommand*\fsize{\dimexpr\f@size pt\relax}
	\newcommand*\lineheight[1]{\fontsize{\fsize}{#1\fsize}\selectfont}
	\ifx\svgwidth\undefined
	\setlength{\unitlength}{424.62546185bp}
	\ifx\svgscale\undefined
	\relax
	\else
	\setlength{\unitlength}{\unitlength * \real{\svgscale}}
	\fi
	\else
	\setlength{\unitlength}{\svgwidth}
	\fi
	\global\let\svgwidth\undefined
	\global\let\svgscale\undefined
	\makeatother
	\begin{picture}(1,1.16139859)
	\lineheight{1}
	\setlength\tabcolsep{0pt}
	\put(0.21665691,1.1171212){\color[rgb]{0,0,0}\makebox(0,0)[lt]{\lineheight{1.25}\smash{\begin{tabular}[t]{l}\textbf{\textsf{Mixed}}\end{tabular}}}}
	\put(-2.38674298,1.37129539){\color[rgb]{0,0,0}\makebox(0,0)[lt]{\begin{minipage}{0.24996292\unitlength}\raggedright \end{minipage}}}
	\put(0.21665691,0.80977627){\color[rgb]{0,0,0}\makebox(0,0)[lt]{\lineheight{1.25}\smash{\begin{tabular}[t]{l}\textbf{\textsf{Layered}}\end{tabular}}}}
	\put(0.21649133,0.50117036){\color[rgb]{0,0,0}\makebox(0,0)[lt]{\lineheight{1.25}\smash{\begin{tabular}[t]{l}\textbf{\textsf{Fractionation}}\end{tabular}}}}
	\put(0.21705708,0.23116924){\color[rgb]{0,0,0}\makebox(0,0)[lt]{\lineheight{1.25}\smash{\begin{tabular}[t]{l}\textbf{\textsf{Water}}\\\textbf{\textsf{Sequestration}}\end{tabular}}}}
	\put(0,0){\includegraphics[width=\unitlength,page=1]{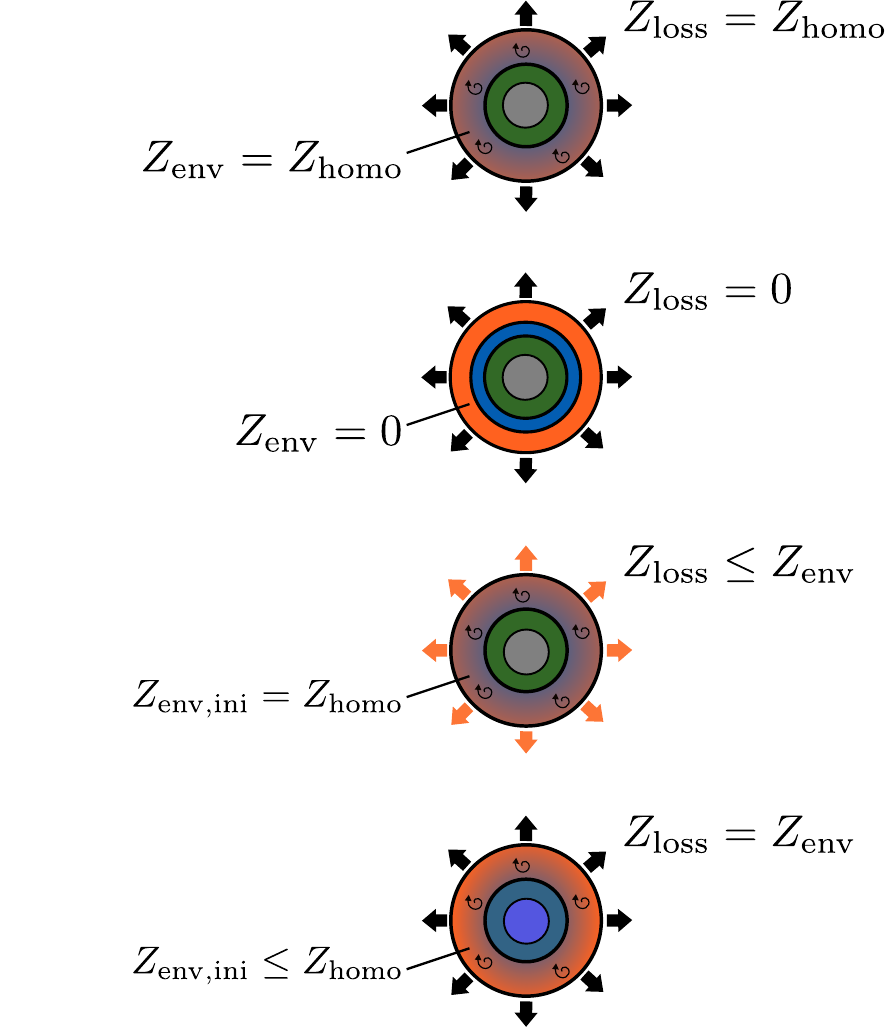}}
	\end{picture}
	\endgroup
	
	\caption{Schematic visualization of the four considered different structure models. Considered layers (inside-out) are: a metallic core (gray); a silicate mantle (green), expected to be in molten form as a magma ocean; in the Layered model a water layer (blue); and at the top, a gaseous envelope (orange). Its metallicity, $Z_{\rm env}$, is in the Mixed and initially in the Fractionation case set to $Z_{\rm homo}$ -- the metallicity resulting from mixing all volatiles with H/He uniformly in the gaseous envelope of the planet. For the Water Sequestration model (bottom), $Z_{\rm env}$ is lower since a part of the accreted volatiles are distributed to the magma ocean and metallic core (indicated by blue coloring of those two layers). The Layered model does not consider volatiles in the envelope. The arrows indicate mass loss and are colored in the Fractionation model to indicate varying metallicities in the lost gas (equal to or smaller than the envelope metallicity).}
	\label{fig:model_schematics}
\end{figure}
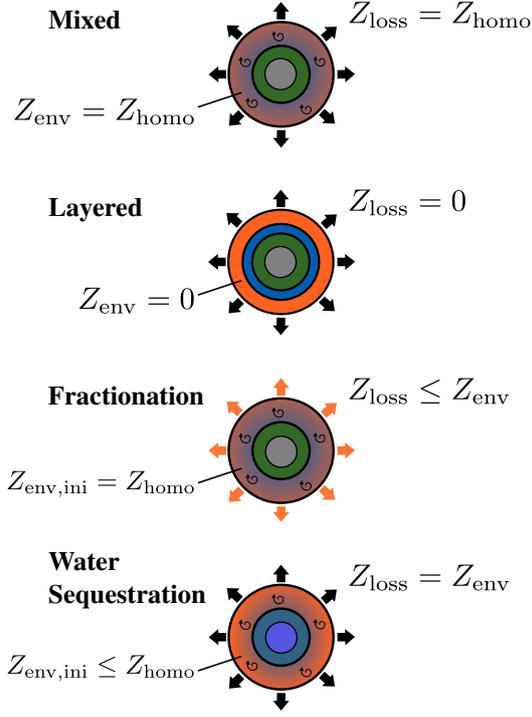

\paragraph{Mixed}
As for the solid part of the planetary structure, we require an equation of state for the envelope matter. For the Mixed model, the envelope is assumed to consist of a perfect mixture of water, hydrogen, and helium. Therefore, the envelope metallicity is set to $Z_{\rm env} = Z_{\rm homo}$. While perfect mixing at all conditions is an extreme assumption, there are clear indications that H/He and water are well miscible at the high temperatures within the runaway greenhouse limit \citep{Vazan2022,Pierrehumbert2023}, which is a safe assumption for the observed population of small transiting exoplanets. The equation of state for water follows \citet{Haldemann2020} while that for H/He follows \citet{Chabrier2019}. To combine them, we assume that Amagat's law of additive volumes holds.

\paragraph{Layered}
In the Layered model, water and H/He are assumed to not mix at all to explore the other fringe case. We consider this model interesting from a theoretical perspective but physically less likely. Further, it resembles the classical onion-like structure assumption often used in the field. In this case, the resolved planetary structure is split into two part, an upper, pure H/He layer on top of the layer including all volatiles, treated as pure water. The uppermost layer follows the general structure outlined above. The pure water layer is solved assuming that the structure is adiabatic (i.e. solving Eqs. \ref{eq:dmdr}, \ref{eq:dPdr}, \ref{eq:dTdr} but using the adiabatic gradient -- Eq. \ref{eq:adgrad} -- in any case in Eq. \ref{eq:dTdr}) and employs the AQUA equation of state to close the system \citep{Haldemann2020}. It uses a fixed 2000 layers and obtains the solution for a given external pressure and temperature, obtained from the previous iteration of the H/He layer, via bisection. The interplay of the two structures is handled by five-fold global iteration per timestep, which is sufficient in phases of slow evolution. Testing different iteration procedures only gave slight differences in the timing (by a few years) of complete loss of H/He, thus we consider this approach acceptable.

After the loss of H/He (i.e. falling below a threshold mass of \SI{8e-7}{\mearth}), we switch for the volatiles to a pure water envelope with the full set of the general treatment but without H/He, thus using pure AQUA EOS. This transition implies for hot planets that we first force the water to be adiabatic, motivated by the pressure of the H/He on top of it, and then transform it to a typically supercritical vapor envelope which is partially radiative. We note that for planets which do not contain volatile ices or H/He after formation the Layered and Mixed models are equivalent.

\paragraph{Fractionation}
The structure and initialization of the Fractionation model is equal to the Mixed model explained above. The difference lies in loosening the assumption that the gas is perfectly mixed also in the photoevaporative wind. Instead, an analytic prescription \citep{Hunten1987,Zahnle1986,Zahnle1990} detailed in Appendix \ref{app:fractionation} (optimistic estimate)\footnote{We use the analytic estimates from Appendix \ref{app:fractionation} and the expressions for $\dot{M}_{\rm H_2O}$ there instead of the output from the tables of \citet{Johnstone2020} (Sect. \ref{sec:photoevap}) for its broader applicability.} is used to determine the metallicity in the photoevaporative mass loss $Z_{\rm loss} = \frac{\dot{M}_{\rm H_2O}}{\dot{M}_{\rm H/He} + \dot{M}_{\rm H_2O}}\leq Z_{\rm env}$. Consequently, this also implies that $Z_{\rm env}$ can evolve, that is, increase, over time.

\paragraph{Water Sequestration}
A novelty in evolutionary calculations is the here presented Water Sequestration model. Instead of mixing the accreted volatile species (treated as water) mass $M_{\rm vol}$ only within the envelope, we maximize the amount of water sequestered in the deep interior by assuming the presence of a molten magma ocean during the whole evolution of the planet. In that case, some amount of volatiles $M_{\rm vol,magma}$ can dissolve into the magma and $M_{\rm vol, core}$ can even dissolve into the metallic core based on the calculations in \citet{Luo2024} \citep[see also][]{Dorn2021}. The envelope volatile mass at the start of our long-term evolution phase is then only $M_{\rm vol, env} = M_{\rm vol} - M_{\rm vol, magma} - M_{\rm vol, core}$ and the envelope metallicity $Z_{\rm env, ini} = M_{\rm vol, env}/(M_{\rm vol, env} + M_{\rm H/He})$. The rationale of the model is to provide an upper limit for the mass of volatile species, treated as water, which could be removed from the envelope mass budget. This has implications for the evolution of the planetary envelope. To provide this upper limit of the effect, we assume a completely molten mantle (see discussion in Sect. \ref{sec:sequestration_discussion}) which can incorporate orders of magnitude more volatile mass compared to a solid mantle \citep{Shah2021}.

\begin{figure}
	\centering
	\includegraphics[width=.9\linewidth]{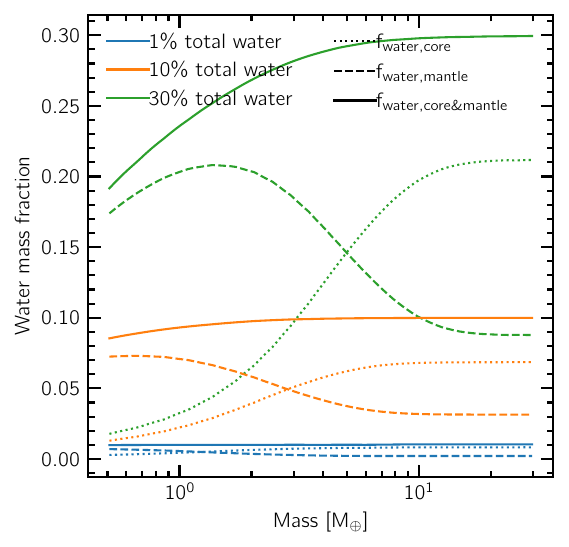}
	\caption{Water mass fractions in core and mantle relative to total mass of the planet. These values provide an upper limit and are used in the Water Sequestration model. The three different colors indicate cases with 1, 10, and 30\,\% of total planetary volatile mass content (including the gaseous envelope) $M_{\rm vol}/M$ while the different line-styles show the two interior reservoirs (dotted for core, dashed for mantle), respectively the total water mass fraction sequestered in the interior (solid).}
	\label{fig:water_fraction_core}
\end{figure}

The partitioning of water between the mantle and surface water layers is modeled using a modified version of Henry’s law which relates the water mass fraction in the magma $M_{\rm vol, magma}/(M_{\rm vol, magma} + M_{\rm Rock})$ to the pressure at the bottom of the gaseous envelope $P_{\rm B}$ (see \citet{Dorn2021} for details).
\begin{equation}
\label{eq:solubility}
M_{\rm vol, magma}/(M_{\rm vol, magma} + M_{\rm Rock}) = \alpha P_{\rm B}^{1/\beta}\,.
\end{equation}
The parameters $\alpha > 0 $ and $\beta > 0$ were fitted to experimental data by \citet{Dorn2021}. Partitioning from the magma ocean to the metallic core (to determine $M_{\rm vol, core}$) follows \citet{Luo2024} who assume iron silicate equilibration at mid-mantle pressures. Their partitioning coefficients depend on the concentration of water in the molten silicates as well as mid-mantle pressures but only marginally on the temperature at a given pressure.

For the coupling to the evolutionary cooling and photoevaporation model, tabulated results from these partitioning and solubility laws and further using the interior structure model detailed in Appendix \ref{app:core_density_effect_water_in_core} were used. The planetary structure for which a solution is found when creating the tables includes core, mantle, and water atmosphere. Thus, a consistent planet structure with water partitioning to core and mantle can be found via iteration for a given total mass and water mass fraction. We note that we did not include an explicit temperature dependency on the water partitioning leading to a weak effect of external temperature via the pressure dependencies. Therefore, the tables were using an exterior temperature of 3000\,K for all models and varying total water mass budgets (from 1\% to 30\%) and planetary masses (0.5 to 30\,M$_\oplus$) were considered. For the equilibration of water between silicates and iron, we use mid-mantle pressures and the iron mass fraction of the dry core was assumed to be 0.244 consistent with the outcome of the equilibrium condensation model used in our formation model \citep{Thiabaud2014}. To construct the tables, the presence of H/He, which would increase pressures and thus solubilities, was excluded for simplicity but supported by the fact that most planets in the considered size regime contain little H/He in mass.

The resulting water mass fractions distributed at initialization to the molten mantle and the metallic core are shown in Fig. \ref{fig:water_fraction_core}. We see a general trend of increasing water budget in the metallic core with increasing total mass. This is expected given the more siderophile nature of water with increasing pressure. The magma ocean water content peaks at intermediate masses. Combined, this results in a moderately increasing mass-dependency for the total fraction of water in the solid interior (solid lines).

During time evolution, the total mass of the planet and the atmospheric pressure can decrease due to photoevaporative mass loss. Therefore, the balance between sequestered water in the magma ocean and envelope (Eq. \ref{eq:solubility}) is re-evaluated after each timestep and $M_{\rm vol, magma}$ is re-determined using the same tables as initially. The water partitioning between envelope and magma ocean is therefore dynamically adapted to the new conditions which leads to outgassing of water from the magma and a gradual enrichment of the envelope with water vapor until $Z_{\rm env} = 1$ is reached. In contrast, we assume that the volatiles trapped in the iron core remain constant over time (similar to outgassed scenarios in \citealp{Rogers2024a}). The same tabulated data can be used because the water reservoir in the magma ocean is not affected by the amount of water in the core but by exterior conditions.

For simplicity, the Water Sequestration model calculates the average density of the volatile-enriched metallic core and magma ocean using the equation of state of \citet{Seager2007} assuming the same separation into iron core, perovskite mantle, and water ice layer as in the Layered model. This implies that for the radius and bulk density of the interior, we do not assume that sequestration to the mantle and core occurred. Instead, the sequestered water (of mass $M_{\rm vol, magma} + M_{\rm vol, core}$) is treated as a high-pressure ice layer on top of the rocky material. A comparison to densities calculated consistently with the structure assumption of the Water Sequestration model (Fig. \ref{fig:model_schematics}) is presented in Appendix \ref{app:core_density_effect_water_in_core}.

\subsubsection{Photoevaporative mass loss}
\label{sec:photoevap}
To model the escape of particles from the upper atmosphere due to high-energy radiation, we use a similar prescription for photoevaporation as in \citet{Affolter2023} and \citet{Burn2024}. We assume that the escape is in the thermally-driven hydrodynamic regime \citep[see][for a recent review]{Gronoff2020}. The total mass loss is a combination of the loss of H/He \citep{Kubyshkina2021} and that of volatile species treated as water \citep{Johnstone2020}. Given the $Z_{\rm env}$, differing between models (see Sect. \ref{sec:structure_and_compo}), the total mass loss rate is given by
\begin{equation}
\dot{M}_{\mathrm{evap}} = Z_{\rm env} \dot{M}_{\mathrm{H}_2\mathrm{O, J}} + (1-Z_{\rm env}) \dot{M}_{\mathrm{H/He, Ku}}\,,
\label{eq:Mdot_total}
\end{equation}
where we use the tabulated values of \citet{Kubyshkina2021} for $ \dot{M}_{\rm H/He, Ku}$. The results are based on the chemical-hydrodynamic models of \citet{Kubyshkina2018}, assume a 15\% heating efficiency of the high-energy radiation, and depend on stellar mass, X-ray and extreme ultraviolet (EUV) fluxes $F_{\rm XEUV}$, planetary masses and radii, as well as equilibrium temperature. To transition from these models applicable for pure H/He planets to water-dominated cases, the results of \citet{Johnstone2020} are used to determine $\dot{M}_{\mathrm{H}_2\mathrm{O, J}}$ in Eq. \eqref{eq:Mdot_total} by adjusting the efficiency parameter $\epsilon$ in the energy-limited mass loss equation
\begin{equation}
	\dot{M}_{\rm esc,EL} = \epsilon \frac{\pi F_{\rm XEUV} R_{\tau=2/3} R_{\rm base}^2}{G M_{\rm tot} K(\xi)}\,,
\label{eq:energy_limited_escape}
\end{equation}
where $F_{\rm XEUV}$ is the received flux at the planet location in either X-ray (dominating early) or EUV wavelengths, $R_{\tau=2/3}$ is the radius of the layer where an optical depth of 2/3 is reached from the inside-out, and $R_{\rm base}$ is the base of the ionization layer. It is found by searching within the resolved planetary structure for the location where an optical depth of one if reached for UV photons \citep{Murray-Clay2009}. The factor $K(\xi) = 1 - \frac{3}{2 \xi} - \frac{1}{2 \xi^3}$, with $\xi = R_{\rm Roche}/R_{\tau=2/3}$ being the ratio of the planets Roche limit to its radius. As in \citet{Burn2024}, the time evolution of $F_{\rm XEUV}$ follows \citep{Mcdonald2019} and we linearly interpolate or extrapolate to lower stellar masses if required. In contrast to \citet{Burn2024}, we reduced the rates of \citet{Johnstone2020} to account for an additional H$_2$O-abundance-dependent cooling effect fitted following \citet{Yoshida2022} as $\dot{M}_{\mathrm{H}_2\mathrm{O, J}} = \dot{M}_{\mathrm{H}_2\mathrm{O, J, non-mod}} \times 1/(0.5 \log_{10}(Z_{\rm env}) + 3)$, where $\dot{M}_{\mathrm{H}_2\mathrm{O, J, non-mod}}$ is the unmodified rate fit to the results of \citet{Johnstone2020}. We note that the precision of this approach is order-of-magnitude only and that the work by \citet{Yoshida2022} was conducted using the XUV spectrum of an M8 star (motivated by TRAPPIST-1), which differs significantly from a Solar-type star and is therefore most appropriate at the low stellar mass end.

\subsection{Observational data and sampling}
\label{sec:sampling}
To compare simulations and observations, we use the collection of well characterized exoplanets from \citet{Parc2024} with the goal of comparing mass-radius relations. For our purposes, we restrict ourselves to planets within 30\,days orbital periods and smaller than 7\,R$_\oplus$. Furthermore we require precisely determined masses and radii with relative errors below 25\% and 8\% respectively. This results in a set of 201 observed planets. As a second exploration, we will relate the Solar-mass population of simulated exoplanet radii to the data from the California-Kepler survey \citep{Fulton2018}. For this second set of data, no precise masses are available but the application of the bias from Kepler can be done in a rigorous fashion using the KOBE package \citet{Mishra2021}.

To allow for a tentative, quantitative comparison also for the sample from \citet{Parc2024}, we apply an estimate for the observational bias and re-sample the synthetic population that is an outcome of the model described above such that the distribution of stellar masses and the right number of observed stars is reproduced. The observational data originates from heterogeneous sources and a human intervention bias is introduced when follow-up characterization observations are scheduled. Nevertheless, a significant fraction of the planets were discovered by TESS \citep{Ricker2015} which is a transit survey. Therefore, we calculate the geometrical transit probability \citep{Petigura2018}
\begin{equation}
p_{\rm trans} = 0.9 \frac{R_\star}{a}
\end{equation}
and an estimate for the probability that a transiting planet is also detected \citep[based on the Kepler mission, as detailed by][]{Fulton2017,Petigura2018}
\begin{equation}
p_{\rm det} = \Phi\left(R_{\mathrm{trans}},\log_{10}(\mu),\sigma\right)\,, 
\end{equation}
where $\Phi$ is the standard cumulative distribution function, $\mu = \SI{1.387}{R_{\oplus}} \times \sqrt{T_{\rm Kepler}/T_{\rm TESS}} \times (P/\SI{100}{days})^{0.19}$, and $\sigma=\SI{0.145}{dex} \times 2$. Here, we use the expected scaling of signal to noise with observation time \citep{Petigura2018}, estimate the mission duration ratio of Kepler to TESS $T_{\rm Kepler}/T_{\rm TESS} = 13.33$, and increase $\sigma$ by a factor of two to account for the heterogeneous sample as well as the variability in TESS observation duration per star (depends on the sector). We visually verify the estimated detection probabilities by comparing them to the results of the detailed analysis of TESS planet detection yields by \citet{Kunimoto2022}.

Most of the models (except the Layered one) rely on the assumption that water mixes with H/He. This is established for planets which are interior to the runaway greenhouse limit where water cannot condense due to too high temperatures. Therefore, we apply a cut at the runaway greenhouse limit which affects planets around different stars differently: The runaway greenhouse limit lies at close orbits similar to the detection thresholds for late M dwarfs only. For the estimate we use either the observed luminosities for the observational data or for synthetic stars, we use stellar luminosities $L_{\star}$ at \SI{5}{Gyr} from the tracks by \citet{Baraffe2015}. The incident flux of shortwave radiation on the planetary surface is on average $I_{\rm ISR} = (1-\alpha_{\rm al}) L_{\star}/( 16 \pi a^2 )$, where $\alpha_{\rm al}$ is the planets' albedo. Assuming that the planet is in thermal equilibrium and for a limiting case assuming no internal heat content, we can equate this to a critical outgoing long-wave radiation limit of $I_{\mathrm{OLR}} = \SI{281}{\watt\per\square\meter}$ from \citet{Boukrouche2021}. The resulting critical distance
\begin{equation}
a_{\rm runaway} = \sqrt{  \frac{ (1-\alpha_{\rm al}) L_{\star}}{ 16 \pi I_{\mathrm{OLR}}  } }\,,
\label{eq:runaway}
\end{equation}
is then a conservative upper limit within which our assumption that the water evaporates and mixes with other constituents holds. To remain conservative, we use a relatively high albedo of $\alpha_{\rm al}= 0.4$ motivated by the results of \citet{Kopparapu2014} for high-pressure envelopes. 

To generate a synthetic sample of planets we sample \SI{10000}{} systems from the original synthetic data weighted by stellar mass chosen such that the stellar mass distribution of the synthetic data matches the observed one \citep[equivalent to][]{Schlecker2022}. We then proceed to draw ``observed'' planets by repeatedly (1000 times a set of 201) sampling planets weighted by their individual $p_{\rm trans} \times p_{\rm det}$ thus allowing for an individual, modeled planet to be present several times in the synthetic comparison sample. We note that this approach ignores possible correlations of the detection probability within a planetary system, and we used discrete stellar mass bins (0.1, 0.3, 0.5, 0.7, and \SI{1.0}{\msol}) inherited from the formation model of \citet{Burn2021}.

Since the observed planets, to which we want to compare our models, are also characterized by radial velocity measurements, we further require for planets to be included in the synthetic comparison data that the semi-amplitude \citep[e.g.][their Eq. 1]{Cumming1999} is larger than \SI{0.3}{\meter\per\second}.

\section{Results}
\label{sec:results}
\subsection{Evolutionary tracks}
\label{sec:evo_track}
\begin{table}
	\caption{Initial conditions for the planetary evolution tracks shown in Fig. \ref{fig:evo_tracks} and discussed in Sect. \ref{sec:evo_track}.}
	\begin{tabular}{r | c | c | c}
		\hline\hline\rule{0pt}{2.6ex}
		& Light & Intermediate & Massive \\
		\hline\rule{0pt}{2.6ex}
		Total mass [M$_\oplus$] & 2.76 & 5.57 & 10.92 \\
		Envelope mass $M_{\rm env}$ [M$_\oplus$] & 1.07 & 3.31 & 6.24  \\
		Core mass $M_{\rm Fe+Si}$ [M$_\oplus$] & 1.69 & 2.26 & 4.68 \\
		Iron mass fraction $f_{\rm iron}$ & 0.24 & 0.24 & 0.24 \\
		Envelope metallicity $Z_{\rm homo}$ & 0.79 & 0.73 & 0.67 \\
		Luminosity $L_{\rm ini}$ [L$_{\rm Jup}$] & 2.81 & 12.29 & 50.50 \\
		\hline
	\end{tabular}
	\tablefoot{The core and envelope masses are for the Mixed composition, for the Layered and Sequestration models the envelope and core masses as well as the metallicity are modified.}
	\label{tab:evo_track_parameters}
\end{table}

\begin{figure}
	\centering
	\includegraphics[width=.9\linewidth]{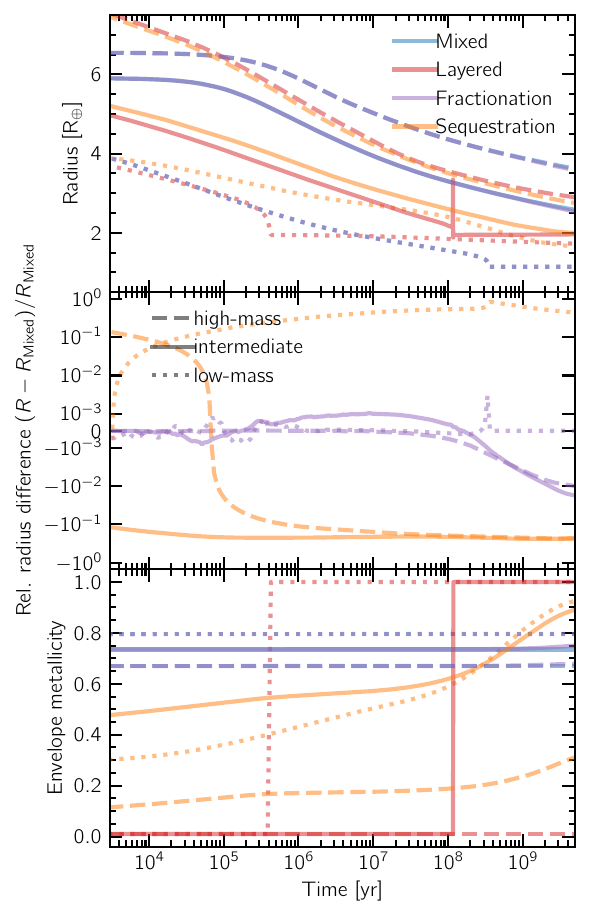}
	\caption{Time evolution of planets using the different models. Top and middle panel show radii and radius difference to the Mixed model; the bottom plot envelope metallicities $Z_{\rm env}$. Time is measured relative to the start of the evolution, here fixed to 3\,Myr. The dashed lines show a more massive sub-Neptune with an initial mass of \SI{10.92}{\mearth}, the solid ones that of an intermediate case (initial mass \SI{5.57}{\mearth}), and the dotted lines represent the evolution of a lighter planet \SI{2.76}{\mearth}. All planets are placed at 0.01\,au around a 0.5\,M$_\odot$ star (see text for further initial conditions). Note that the Fractionation model results overlap with the Mixed model result in the top panel. For this reason, the middle plot shows the relative radius difference to the equal-initial mass Mixed model case.}
	\label{fig:evo_tracks}
\end{figure}
In order to understand the dependencies and behavior of the different model assumptions, it is illustrative to visualize the time evolution of three different planetary mass cases under the four different model assumptions. This covers qualitatively the different possibilities for the final states of initially water-rich planets. Here, we initialized the planets by using a total mass -- luminosity relation of \citet{Linder2019}, and H/He and water fractions typical for the formation output. The parameters are tabulated in Table \ref{tab:evo_track_parameters}. Furthermore, the planets are placed after \SI{3e6}{\year} at the same orbit with semi-major axis of 0.01\,au around a 0.5\,M$_\odot$ star.

At time zero of the model, the Water Sequestration and Layered models distribute a fraction or all of the water contained in the envelope $Z_{\rm homo} M_{\rm env}$ either to the core or it is layered on top of the core. The gaseous envelope is therefore depleted in heavy particles and adjusts itself to a larger radius. As soon as the simulation starts evolving in time, the larger radius and H/He-dominated composition lead to a higher photoevaporative mass loss rate. The latter is caused by up to an order of magnitude larger loss-rates for H/He than H$_2$O for equal mass and radius planets when using the prescriptions outlined in Sect. \ref{sec:photoevap} based on \citet{Kubyshkina2021} and \citet{Johnstone2020}. For further discussion on the limitations of the photoevaporation model, we refer to Sect. \ref{sec:discussion_photoevaporation}. Thus, the planet loses mass and consequently shrinks in size more rapidly than the cases with a Mixed or Fractionation assumption. This can be observed for the intermediate and massive planet and the effect is slightly more pronounced in the Layered compared to the Water Sequestration scenario as all water, instead of just a part, corresponding to the possible uptake of the core and magma ocean, is removed from the envelope.

Depending on the initial mass and structure assumption, the gaseous envelope can be lost completely. For the Layered model, the $Z_{\rm env}=0$ envelopes are removed in the intermediate and low mass cases and replaced by pure water $Z_{\rm env} = 1$ envelopes from the layer below. The intermediate-mass case goes through an inflation phase shortly before the H/He mass is completely lost just after \SI{100}{Myr}. This is not completely physically correct as the AQUA layer is assumed to remain fully adiabatic in our simplified treatment during the evolution phase. In reality, it should become partially radiative. The outcome is nevertheless as expected such that the H/He is fully lost and a new consistent hydrostatic envelope structure (incl. a radiative layer) made of pure water $Z_{\rm env} = 1$ is initialized.

Similarly, for the Water Sequestration model, the most massive planet retains its gaseous envelope while the intermediate and light planets transition to a secondary envelope stage via the equilibration of the interior with the atmosphere, that is, due to gradual outgassing of water. It leads to an increase in $Z_{\rm env}$ as can be seen in the bottom plot. As mentioned in the model description and will be discussed in Sect. \ref{sec:sequestration_discussion}, this approach is relatively crude and could be improved in the future. For lower mass planets (here the intermediate and low-mass case), the resulting atmosphere is almost entirely made of heavy elements $Z_{\rm env}>0.8$. Although it could be removed on timescale comparable to the lifetime of the system, the replenishment from the much larger magma ocean reservoir typically sustains it for planets with $M>2\,M_\oplus$ \citep{Kite2021}. This result also already hints at a bifurcation of $Z_{\rm env}$ with either low values, where hydrogen and oxygen are still stored in the core, or high $Z_{\rm env}$ for planets which underwent efficient mass loss and significant outgassing.

Due to the faster loss of H/He in the initial stage, the Layered and Water Sequestration model yield planets with lower radii compared to the Mixed model, except for cases where outgassing was significant. This difference would last up to the point at which the Mixed model planet loses its atmosphere completely. In the cases shown in Fig. \ref{fig:evo_tracks}, this occurs only for the light planet after 400\,Myr. Since this planet in the Mixed model will have lost not only H/He but also its complete water budget, the planet ends as a smaller rocky core compared to the Water Sequestration model where an outgassed atmosphere remains and even more so compared to the Layered model result where a thicker steam atmosphere is present.

The Fractionation model differs less than a percent in radius from the Mixed model. There are hints of weakly fractionating mass-loss at the end of the lifetime of the atmosphere on the light planet (400\,Myr, dotted purple line) where the atmosphere is removed slightly faster in the Mixed model case. For the more massive planets, the effect of fractionation is to marginally increase $Z_{\rm env}$ over time which leads to percent-level smaller radii due to the slightly increased mean molecular weight.

\subsection{The mass-radius-period distribution in the steam and stone interpretation}
\begin{figure*}
	\centering
	\includegraphics[width=\linewidth]{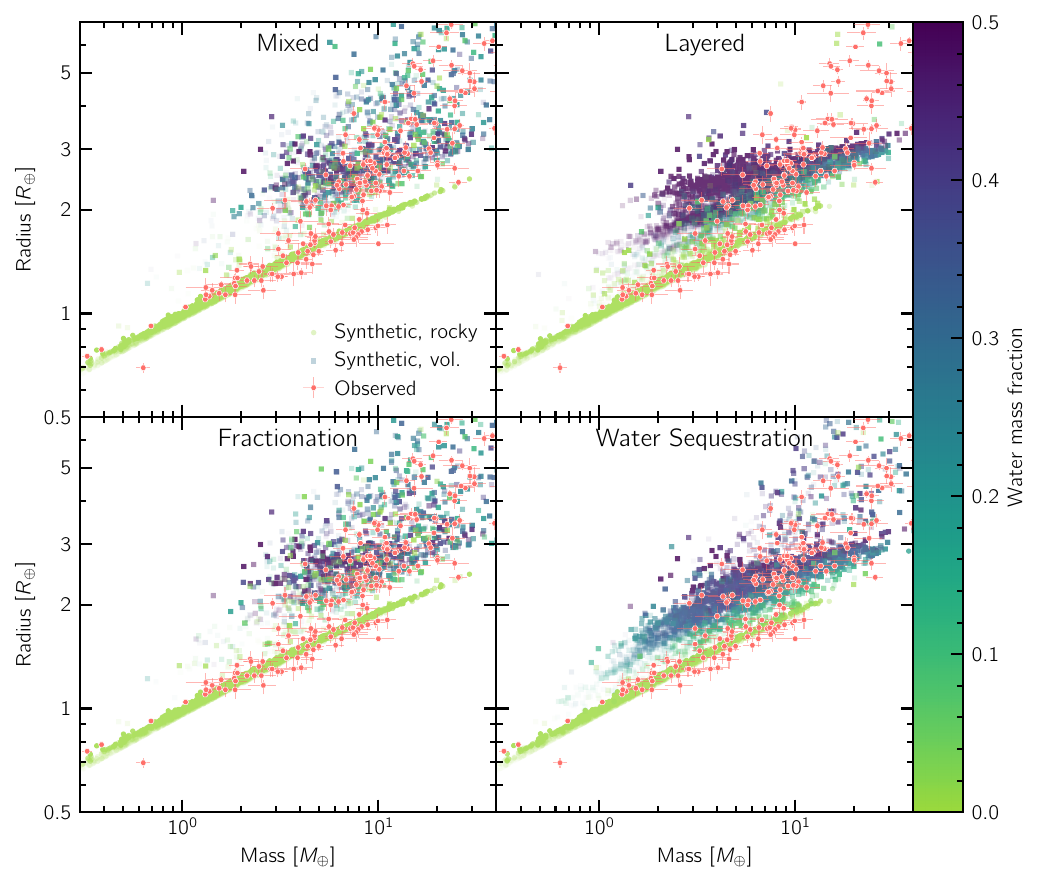}
	\caption{Mass versus radius diagram of observed and synthetic planets. The four panels show the resulting planets using different model variations contrasted against the observational data from \citet{Parc2024}. The synthetic data for all models is sampled according to the stellar mass distribution of the observations and an estimate of the detection and transit probability is applied. More saturated dots imply that the planet was sampled multiple times due to being more likely to be detected and their color is given by the planets' water mass fraction.}
	\label{fig:m_R_all}
\end{figure*}

\subsubsection{Mixed model}
We start by comparing the mass-radius relation of the synthetic planets using the nominal, mixed envelope assumption against the sample of observed planets in the top, left panel of Fig. \ref{fig:m_R_all}. The synthetic distribution consists of volatile-rich planets with significant scatter in mass-radius relation distinct from a population of higher-density, rocky planets with a tightly correlated mass-radius relation.

The rocky population seems to be in agreement with the observational data although there might be too many massive rocky bodies. The upper end of the rocky body distribution will be discussed quantitatively in Sect. \ref{sec:fits}.

The volatile-rich population of the Mixed model has a decreasing number density with increasing radius which is more clearly seen if estimates of the occurrence probability density are added (Appendix \ref{app:supp_figures}). The contours in Fig. \ref{fig:m_R_zhomo_kde} show this information and make apparent that both the observed and synthetic planets are more likely to occur at around 2.7\,R$_\oplus$ in contrast to larger radii. The synthetic planets however populate a parameter space at low masses and large radii where few observations are found. This regime is sensitive to the observational bias since the radial velocity semi-amplitudes are low. With increasing radius, there is no clear water mass fraction correlation visible in Fig. \ref{fig:m_R_all} but we further explored envelope metallicities, which do show a decreasing trend with increasing radius.

\begin{figure*}
	\centering
	\includegraphics[width=\linewidth]{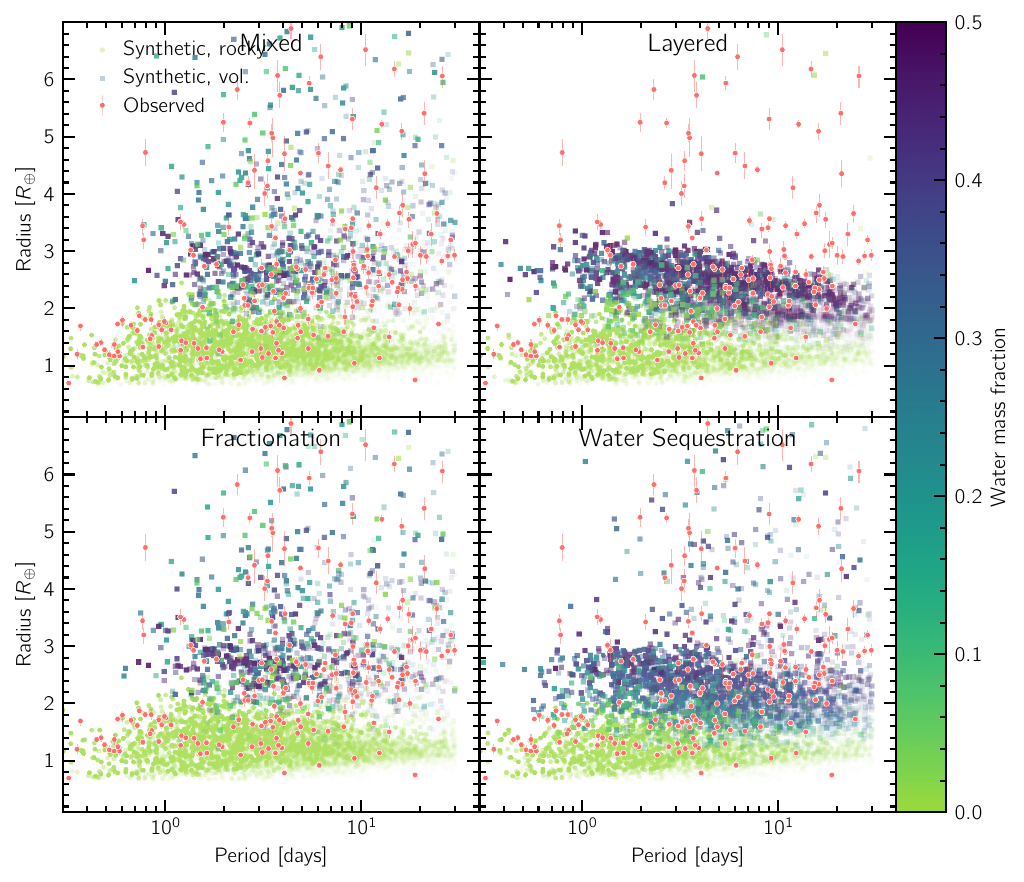}
	\caption{Orbital period against planetary radii of observed and synthetic planets. The samples and coloring are the same as in Fig.  \ref{fig:m_R_all}.}
	\label{fig:P_R_all}
\end{figure*}

In addition to masses and radii, it is insightful to also compare orbital periods of the planets. Here, observations give very precise measurements which is why we omitted their errors from Fig. \ref{fig:P_R_all}. For the Mixed model shown in the upper left panel, we can see an under-density of planets corresponding to the observed radius valley. It separates in our model high-density planets from low-density planets. The Mixed model output agrees well with the observational data in period-radius space. The observation that the innermost planets above the radius valley, the sub-Neptunes, are located at longer orbital periods than the innermost rocky planets is recovered. 

To summarize, we qualitatively compared the Mixed model output against stellar-mass independent observational data in terms of masses, radii, and orbital periods. From this, we cannot discard the Mixed model and reasonably reproduce the observations.

\subsubsection{Layered model}
The Mixed model results can be contrasted with those from the Layered model. In this case, two sharp transitions can be made out in the top right panel in Fig. \ref{fig:m_R_all}. H/He-rich planets differ fundamentally from the water worlds as mixing between the two components is not allowed. Therefore, H/He can also be more easily lost to photoevaporation as loss of H/He is more efficient and H/He-rich planets are initially larger in radius if it is not mixed with heavier species as seen in Fig. \ref{fig:evo_tracks} and discussed in Sect. \ref{sec:evo_track}.

Since the H/He loss is efficient, this leaves a large population of planets with pure water envelopes which is then protected from photoevaporative loss due to both the small radii caused by the higher atmospheric mean molecular weight compared to mixtures and the lower escape rates of pure H$_2$O. Therefore, fewer planets populate the low-density parameter space at larger masses $M>5\,M_{\oplus}$. Instead, many planets are located in the pure water world regime at around two R$_\oplus$.

The lowest-mass volatile-rich planet of each population will be discussed in detail Sect. \ref{sec:fits}, but already from qualitative comparison, the Layered model can be considered a worse match to observations in terms of the volatile-rich planet distribution with too many low-mass planets and too few higher-mass, lower density planets. A similar conclusion can be drawn from a comparison in period-radius space (Fig. \ref{fig:P_R_all}). 

A noteworthy feature is that, in the Layered model, fewer planets populate the massive super-Earth regime. The pathway to form those planets requires stripping of the volatile elements by photoevaporation which is inefficient at those high masses if no light elements are present to decrease the overall density -- as that is the case in the second stage of the Layered model evolution.

\subsubsection{Fractionation model}
The results of the Fractionation model can be summarized briefly: the effect of fractionating mass-loss has negligible consequences for the bulk planet properties using the considered initial conditions. Therefore, the results of the Fractionation model are similar to that of the Mixed model and we will in the following omit a separate discussion for the Fractionation model results. This result is interpreted in Sect. \ref{sec:fractionation_interpretation}. 

\subsubsection{Water Sequestration model}
\label{sec:results_seq}
The results of the Water Sequestration model can be broadly placed between the Layered and the Fractionation model. A part of the water content, usually the majority by mass, separates from the gaseous envelope and is therefore not lost to photoevaporation. From Fig. \ref{fig:m_R_all}, we see that down to a mass of 1\,M$_\oplus$, large water reservoirs can remain on observable planets. At the end of the evolution (see also Sect. \ref{sec:evo_track}), this water is present in the form of oxygen and hydrogen in the metallic core, in the magma ocean, as well as outgassed as steam atmosphere. Planets in this state are tightly correlated in mass-radius while planets with primordial H/He content are distributed to larger radii. The shape of the mass-radius distribution is similar to the output of the Layered model, however, the planets with steam atmospheres are significantly smaller than in that model. This is due to only a fraction of the initial water content being present in gaseous form and reduced total water mass compared to the Layered model.

Compared to observations, the extension of water-rich planets to masses below 3\,M$_{\oplus}$ is not observed (see also Sect. \ref{sec:fits}). Furthermore, from Fig. \ref{fig:P_R_all}, we find a qualitatively worse match to observations with a too narrow radius valley at short orbital periods for the Sequestration model compared to the Mixed one. We refrain from further interpretation of the differences because the observed and modeled distributions do not match well in period space in general which is likely a shortcoming of the formation model and not necessarily related to the long-term evolution.

Despite these differences, the mass-radius distribution of planets with masses above 3\,M$_{\oplus}$, including larger sub-Neptunes, is well-matched in the Water Sequestration model. In particular, no planets with too low densities exist in the sub-Neptune regime while such planets existed in the Mixed model output. The outcome of the model differs in this regime because the same planet (similar to the high-mass case shown in Fig. \ref{fig:evo_tracks}) initially has a lower $Z_{\rm env}$. The even lower density planets than the Mixed case then undergo significant mass loss before outgassing of metals enhances their $Z_{\rm env}$. Combined, this leads to higher density final planets. Furthermore, the initial $Z_{\rm env}$ is more uniform across planets as it is regulated by the core and mantle reservoir reducing the scatter in the distribution.

Lastly, no massive bare rocky planets are predicted in agreement with observations. This indicates that lowering the water fraction in the gaseous envelope led to a better match compared to observations for planets with $M>3\,$M$_{\oplus}$. This is also the regime where the sequestration scenario is most appropriate due to higher accretion energies and thus hotter cores which are more likely to be completely molten.

\subsection{The radius valley as a function of stellar mass}
\label{sec:st_mass_dependency}
\begin{figure}
	\centering
	\includegraphics[width=\linewidth, trim=0 33 0 7, clip]{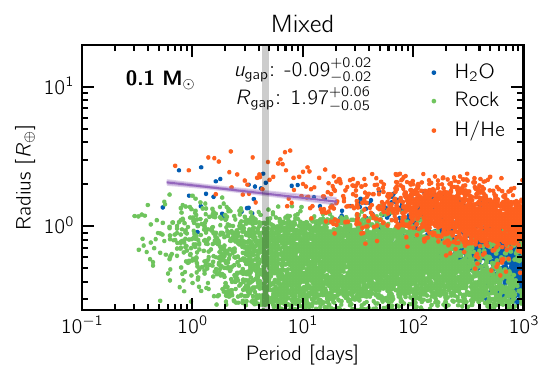}
	\includegraphics[width=\linewidth, trim=0 33 0 7, clip]{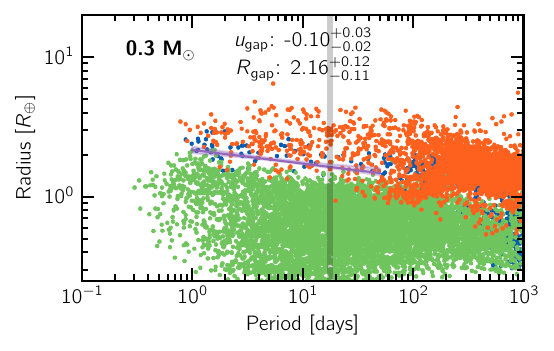}
	\includegraphics[width=\linewidth, trim=0 33 0 7, clip]{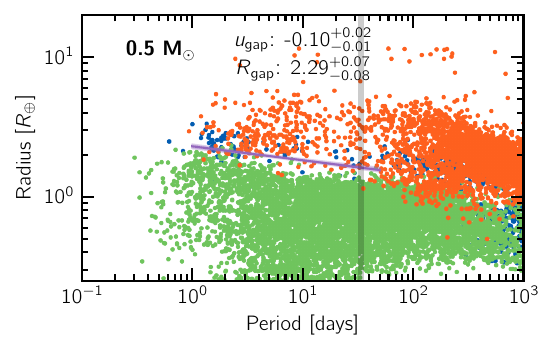}
	\includegraphics[width=\linewidth, trim=0 33 0 7, clip]{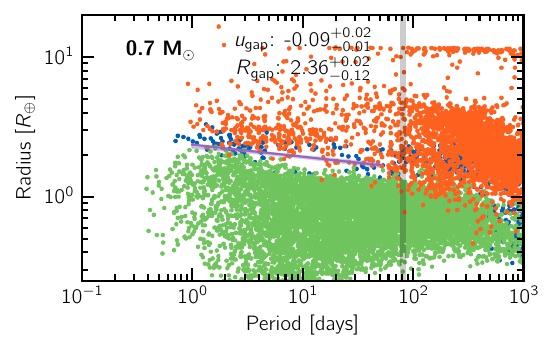}
	\includegraphics[width=\linewidth, trim=0 7 0 7, clip]{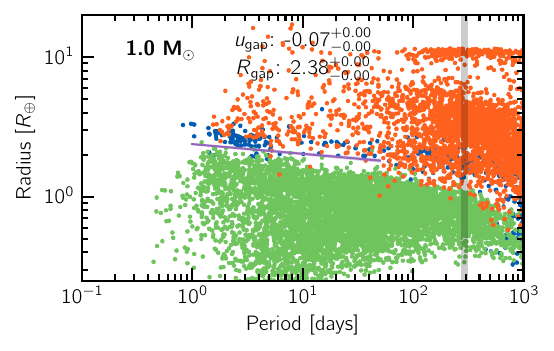}
	\caption{Period versus radius of unbiased, synthetic planets using the Mixed model around stars of different masses. For each stellar mass (top left), a gap in the data was fitted (\texttt{gapfit}, \citealp{Loyd2020}). The gray line marks the runaway limit $a_{\rm runaway}$ (Eq. \ref{eq:runaway}).}
	\label{fig:period_radius_Mstars}
\end{figure}

\begin{figure}
	\centering
	\includegraphics[width=\linewidth]{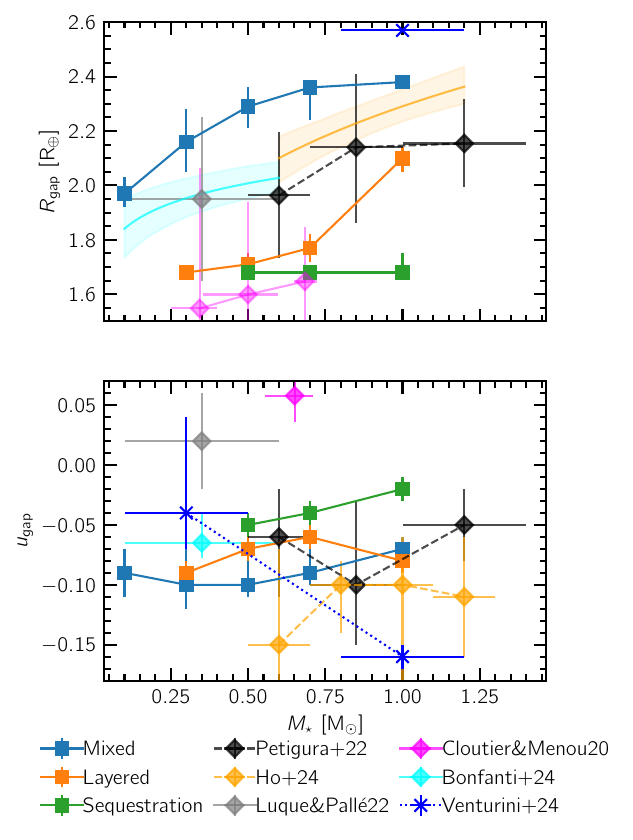}

	\caption{Comparison of derived loci $R_{\rm gap}$ and slopes with orbital period $u_{\rm gap}$ for the radius valley. Observational data from \citet{Petigura2022,Ho2024,Luque2022,Cloutier2020,Bonfanti2024} is shown with diamonds and errors in x-direction indicate the range of stellar masses included. From this work, the Mixed (light blue squares), Layered (red squares), and Water Sequestration (green) model output is included here since the Fractionation model gives similar results to the Mixed case. \citet{Venturini2024} (dark blue) did not obtain an estimate of $R_{\rm gap}$ for their sample of modeled planets around low stellar mass. The results of \citet{Cloutier2020} and \citet{Petigura2022} are based on occurrences instead of detections. For \citet{Petigura2022} the values for $R_{\rm gap}$ were shifted to 1\,day orbital period (incl. Gaussian error propagation). For the results from \citet{Cloutier2020}, we report their values which were obtained from marginalizing over orbital periods. The values are only comparable to the other derivations because their derived $u_{\rm gap}$ is close to zero. For \citet[][green line and region, $M_{\rm star}<0.6\,$M$_{\odot}$]{Bonfanti2024} and \citet[][orange, $0.6$M$_{\odot} \apprle M_{\rm star}\apprle 1.2\,$M$_{\odot}$]{Ho2024}, we show their fitted continuous function for $R_{\rm gap}(M_\star)$ with shaded regions marking 16 to 84\% confidence intervals.}
	\label{fig:rgap_ugap_comparison}
\end{figure}

Here, we investigate the dependency of the emerging radius valley on stellar mass. In Fig. \ref{fig:period_radius_Mstars}, we show for the Mixed model the full synthetic data without an observational bias applied for different discrete stellar masses. Since the previously introduced bias is applicable to planets characterized in mass and radius, we opt to not use an estimate of the bias in this radius comparison (irrespective of planetary masses) to investigate the actual trends predicted by the model. Complementarily, we will discuss the radius valley trends using the populations around Solar mass stars with applied Kepler observational bias in Sect. \ref{sec:radius_valley_sol}. To support the discussion here, we also show a summary of fits done for observed or synthetic data in Fig. \ref{fig:rgap_ugap_comparison} and mass-radius relations resulting from the Mixed model for the different stellar mass bins in Fig. \ref{fig:m_r_stmasses_mixed}. The results of the other models are presented in the same fashion in Appendix \ref{app:supp_figures}. In each panel of Figs. \ref{fig:period_radius_Mstars}, \ref{fig:period_radius_Mstars_layered}, and \ref{fig:period_radius_Mstars_water_in_core}, \texttt{gapfit} \citep{Loyd2020} was used with bootstrapping to infer a gap in the region of the observed gap, that is, using planets from 1 day to 200 days orbital period for stellar masses larger than 0.1\,M$_\odot$ and from 0.6 to 20 days for the 0.1\,M$_\odot$ case. The gap is fit using a power-law
\begin{equation}
R_{\rm gap}(P) = R_{\rm gap} (P/1\,\mathrm{day})^{u_{\rm gap}} \,\mathrm{R}_{\oplus} \,.
\end{equation}
The search for a gap is performed within 0.15 dex of $R_{\rm gap}= 2\,\mathrm{R}_{\oplus}$, with an initial slope guess of $u_{\rm gap}=-0.1$, and a Kernel width of 0.15 dex.

The synthetic data at low stellar mass for the Mixed model shows a similar gap location at \SI{\sim1.9}{\rearth} in agreement with \citet{Luque2022}. We can see a trend with stellar mass: the locus of the valley at \SI{1}{\day} orbital period shifts upward for late M dwarfs before converging to  $R_{\rm gap}=2.38$ for stellar masses larger than \SI{0.5}{\msol}. This conclusion also holds when analyzing the populations as a function of integrated, received EUV or X-ray flux instead of orbital period. This kind of threshold stellar mass is similar to the mass at which the small planet population no longer increases significantly in mass, as discussed in \citet{Burn2021}. In that work, we attributed this to the fact that growth of small planets is no longer limited by the amount of solids in the disk at stellar masses larger than \SI{0.5}{\msol}. We might recover this here, since growth determines the initial mass and thus indirectly the radius of the Super-Earth population.

The observed radius valley indicates a dependency on stellar mass around FGK-type stars based on Kepler and K2 data \citep{Berger2020a,Petigura2022,Ho2023}. The reported logarithmic slopes of the radius valley location with stellar mass $d \log_{10} R_{\rm gap}/d\log_{10} M_{\star}$ range from 0.1 to 0.4 with likely values around 0.2. However, for M stars, \citet{Luque2022} report a slope consistent with no stellar mass dependency \SI{0.08\pm0.12}{}. Although it is likely not the relevant functional form, we fit a power-law to the inferred radius valley locations for the five presented stellar mass bins at 1\,day orbital period and obtain $d\log_{10} R_{\rm gap}/d\log_{10} M_{\star} = 0.08$, which agrees with the observed trend for M dwarfs \citep{Luque2022} but is flatter than found for FGK stars where values around 0.2 are reported \citep{Petigura2022,Ho2023}. From Fig. \ref{fig:rgap_ugap_comparison}, we see that also the data by \citet{Petigura2022} shows a flattening of $d\log_{10} R_{\rm gap}/d\log_{10} M_{\star}$ for the two higher stellar mass bins.

While the locus of the valley increases, our inferred slopes with orbital period hints at a flatter dependency with increasing stellar mass. The values range from $d \log_{10} R/d\log_{10} P = u_{\rm gap} = -0.10$ at \SI{0.3}{\msol} to -0.07 for Solar mass stars. This is comparable to other theoretical works relying on photoevaporation predictions from hydrodynamic models instead of simple energy- or recombination-limited escape models \citep[the latter resulting in steeper slopes][]{Mordasini2020,Affolter2023}. Although the error from the bootstrap method is small, we expect that the flattening is not based on enough statistical data from the model to confirm it as a significant trend. The flattening could be influenced from below, by growing more massive super-Earths at larger orbital periods around more massive stars, or from the top by less efficient photoevaporation and thus smaller radii of volatile-rich planets at fixed orbital period around lower-mass stars. Our results remain for all stellar masses in the regime of negative slopes consistent with photoevaporation being the dominant effect shaping the valley. The compilation of observational $u_{\rm gap}$ (Fig. \ref{fig:rgap_ugap_comparison}) does not show a significant trend. However, the theoretical work of \citet{Venturini2024} predicts a steeper slope around more massive stars which was not recovered here.

\citet{Ho2024} and \citet{Bonfanti2024} gave indications for a more populated, radius valley around low-mass stars. Similarly, \citet{Cloutier2020} already noted a narrowing of the radius valley towards lower stellar masses. Apart from the 0.1\,M$_\odot$ case where the statistics becomes poor, we find a similar trend in Fig. \ref{fig:period_radius_Mstars}. This trend was also recovered and discussed in \citet{Venturini2024}. From the detailed discussion of mass-radius relations (Sect. \ref{sec:fits}), we interpret part of it being linked to the low-mass tail of the volatile distribution extending into the radius valley (but not the density valley, see \citealp{Luque2022,Venturini2024}). We note that the M star sample probes the planetary population at lower irradiations where condensation could occur, but we excluded this region when we plot biased populations. We leave a more detailed analysis of the pollution of the valley to future works.

In addition to the data, we show an estimate of the runaway greenhouse limit (Sect. \ref{sec:sampling}) as gray line in Fig. \ref{fig:period_radius_Mstars}. The fact that for low-mass stars, this limit is not necessarily outside the regime of interest further motivates the investigation of a model which does not assume that water and H/He mix, that is, the Layered model. Thus, in Fig. \ref{fig:period_radius_Mstars_layered}, we show the resulting unbiased planetary population with the previously introduced compositional color code using the Layered model. The most striking effect is that the population of H/He-free sub-Neptunes is more numerous and extends to longer orbital periods than in the Mixed model. If the assumption that no water enters the low-density envelope is made, also planets at larger orbital periods can lose their H/He content. This also leads to a more abrupt transition from H/He free to H/He-rich planets at a threshold radius for a given orbital period. Therefore, a second radius valley emerges between H/He-rich and H/He-free planets at \SIrange{3}{4}{\rearth}. So far and to our knowledge, no reports on such a valley have been published. However, a visual indication of such an under-density can be made out in the data of \citet{Ho2023} for Solar mass stars, their Fig. 1, although it does not seem to be of high enough significance. We further note that in the Solar mass data, the observable region is well within the runaway greenhouse limit where we expect mixing.

\begin{figure}
	\centering
	{\textsf{Mixed}}
	\includegraphics[width=.9\linewidth]{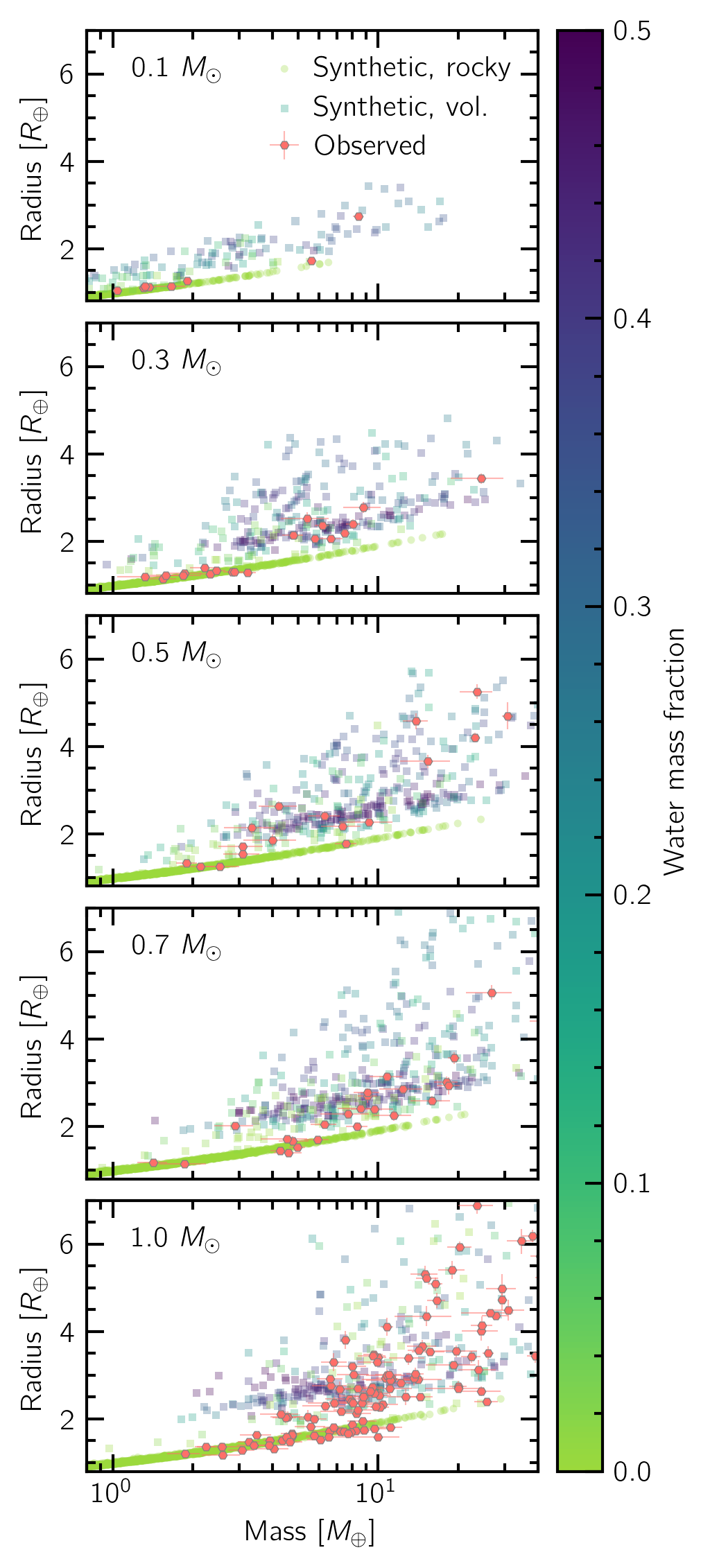}
	\caption{Masses against radii for observed and synthetic planets using the Mixed model for various stellar masses. In contrast to Fig. \ref{fig:m_R_all}, no observational bias is applied. The selection is only restricted to planets within 30\,days, respectively the runaway greenhouse limit (for 0.1 and 0.3\,M$_\odot$). The synthetic data is calculated at the stellar mass indicated at the top left; observational data is grouped by restricting stellar masses \citep[as tabulated by][see references therein]{Parc2024} to the nearest corresponding reference stellar mass.
	}
	\label{fig:m_r_stmasses_mixed}
\end{figure}

For the Layered model, the algorithm could determine a radius valley for all stellar masses except 0.1\,M$_\odot$. The determined valley lies between rocky planets and water worlds. It is located at lower radii compared to observations (Fig. \ref{fig:rgap_ugap_comparison}) because pure water envelopes instead of mixed-composition ones are less prone to being removed by photoevaporation \citep[see e.g.,][for a parameter study]{Mordasini2020}. However, the distance to the observed valley is comparable to the Mixed model. This seems to naively suggest that reality lies between the modeled assumptions.

For the Water Sequestration model, the algorithm also successfully fitted a radius valley for the stellar masses above 0.3\,M$_{\odot}$. In contrast to the previous model results, the core-regulated results do not depend on stellar mass with equal $R_{\rm gap}$ resulting for the different stellar masses. Furthermore, $u_{\rm gap}$ is flatter than in the models driven by photoevaporation. Visual inspection of the results in Fig. \ref{fig:period_radius_Mstars_water_in_core} indicates a more populated radius valley related to the water-rich sub-Neptune branch extending to lower planetary masses and radii (see Fig. \ref{fig:m_R_all}). We also caution that the radius valley in this model emerges because planets with water in their interior efficiently outgas and instantaneously establish an atmosphere in equilibrium with the interior. In a more realistic scenario, there would be a timescale associated to outgassing which would not at all orbital periods be smaller than that of the evaporative loss timescale, thus the radius valley location, which still separates in this model the completely dry from water-rich planets with small outgassed atmospheres, would change or might not emerge at all.

For the Water Sequestration model, no radius valley can be fitted conclusively around the two lowest stellar mass bins. Although there is a transition in density space, the existence of steam worlds with low mass ($\sim2\,$M$_{\oplus}$) makes this density valley undetectable in radius space. This can also be seen when comparing the mass-radius relations shown in Fig. \ref{fig:m_r_stmasses_mixed} with Fig. \ref{fig:m_r_stmasses_waterincore} and again supports the aforementioned interpretations of the radius valley in \citet{Venturini2024} and \citet{Luque2022} although here with a stronger overlap due to the low-mass volatile-rich planets.

\subsection{The radius valley as seen by Kepler}
\label{sec:radius_valley_sol}
\begin{figure}
	\centering
	\includegraphics[width=.9\linewidth]{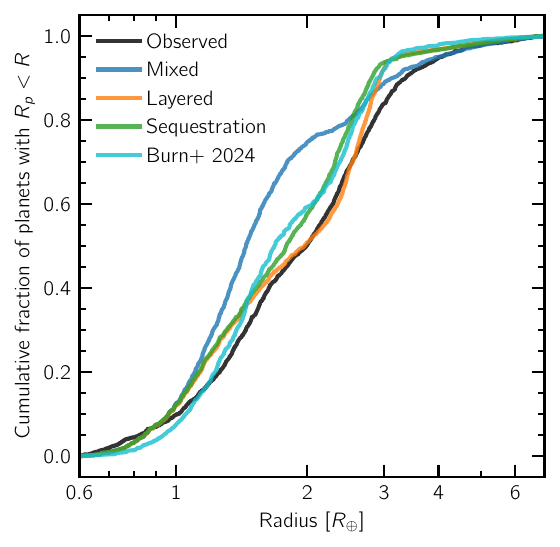}
	\caption{Cumulative distribution of synthetic planetary radii compared to Kepler data. The synthetic data is obtained after modeling the observability with Kepler using KOBE \citep{Mishra2021}. In addition to the datasets obtained here, we show the results of \citet{Burn2024}. Only planets with radii ranging from 0.6 to 7\,R$_\oplus$ are included. The observed distribution is taken from \citet{Fulton2018}.}
	\label{fig:biasedcumulativeobservedmixedlayeredfractionationburn-2024a}
\end{figure}

The ability of the different models to reproduce the observed radius valley can also be tested against the Kepler dataset. As mentioned in Sect. \ref{sec:sampling}, we constructed a set of synthetic systems as the Kepler spacecraft would observe it. Given by the sensitivity of Kepler, we restrict the selection of simulations to those around Solar mass stars. In Fig. \ref{fig:biasedcumulativeobservedmixedlayeredfractionationburn-2024a}, the resulting cumulative distributions of the synthetic planetary radii can be compared to the observed ones. We further added the distribution by \citet{Burn2024} since several assumptions on the initial conditions were changed although the Mixed model is physically similar to the model used in that work.

We find a reduced number of sub-Neptunes in the Mixed model compared to \citet{Burn2024} in disagreement with observations. The high-radius tail of sub-Neptunes, that is, the radius cliff, is best matched in the Mixed model. The radius valley location, also discussed in Sect. \ref{sec:st_mass_dependency} where a radius valley location at 2.4\,R$_{\oplus}$ was found, is shifted to too large radii in the Mixed model. This implies that the sub-Neptunes in the Mixed model lose their envelopes too rapidly \citep[see e.g.][for a parameter study on radius valley location and slope]{Mordasini2020}. However, we caution that the sub-Neptune to rocky planet ratio is sensitive to uncertain formation model assumptions (production of rocky planetesimals and planets, see also discussion in Sect. \ref{sec:discussion_formation_parameters}), as well as to the application of the observational bias. The locus of the radius valley is less sensitive to these considerations.

From Fig. \ref{fig:biasedcumulativeobservedmixedlayeredfractionationburn-2024a}, we find that the Layered model reproduces better the observed radius valley location, although shifted to too low radii for lower stellar masses (see Fig. \ref{fig:rgap_ugap_comparison}). Furthermore, it has a more favorable sub-Neptune to rocky planet ratio compared to the Mixed model. From this comparison on planetary radii only, one might conclude that it should be favored over the Mixed model. In contrast, the Water Sequestration model lacks the distinct radius valley feature in the cumulative distribution.

\section{Discussion}
\label{sec:discussion}
\subsection{Key quantities for comparison of mass-radius relations}
\label{sec:fits}
To better compare the models and observational data, we introduce here four quantities which are not expected to be heavily influenced by observational biases and should relate to different, stellar-mass-dependent processes shaping the period-mass-radius distribution of small planets.
\begin{itemize}
	\item The upper-mass end of the rocky planet distribution $M_{\rm rock,max}$. To not rely on outliers, this is measured as the 90th quantile of the rocky planet mass distribution. Rocky planets are identified by having a normalized density of at least 0.65 times that of an equal-mass Earth composition planet \citep[using the relation by][]{Zeng2016}.
	
	It should be either determined by the most massive planet which can still lose its envelope or the most massive planet which can form in the inner system. If the former is the case, it should be sensitive to the shortest orbital period planets where photoevaporation is strongest. Since transit and radial-velocity methods are most sensitive at short orbital periods, we expect no influence from observational biases.
	
	\item The lower-mass end of the volatile-rich planet distribution $M_{\rm vol, min}$. Analogous to $M_{\rm rock,max}$, the 10th quantile of all other planets but the rocky ones is measured.
	
	Since all planets are expected to form with gaseous envelopes, this quantity is determined by the lightest planet which can retain its envelope. Again, this should be sensitive to orbital periods and is therefore influenced to some degree by the observational bias (the outermost planets which can be detected). However, it is nevertheless a useful property for inter-comparison of models and stellar-mass dependent trends. Alternatively, if stripping of (water-rich) envelopes is inefficient, this point could also be set by the lowest-mass planet which is volatile-rich and observable. Practically, since planets beyond the water iceline are not observable, this would give direct access to the lowest-mass planet which migrated from beyond the water iceline.
	
	\item The volatile-planet mass-radius relation slope $m_{\rm vol,\,M-R}$. This can be fitted using least-squares regression as implemented in the \texttt{scipy} python package \citep{scipy}. By using a single slope, we assume that the mass-radius relation is of the form $R(M) = R_0/R_{\oplus} (M/M_{\oplus})^{m_{\rm vol,\,M-R}}$. To estimate uncertainties, we use a bootstrapping approach for the observed data: for 1000 repetitions, an equal-sized distribution of planets is drawn (with replacement) from the observations. In addition, we randomize the observed values by sampling from a skewed normal distribution based on the errors in mass and radius space. For synthetic data, we obtain comparable error estimates by drawing 1000 sub-samples of planets with the same number of points as the observed data, using a Jackknife approach (i.e. without replacement).
	
	Physically, the slope $m_{\rm vol,\,M-R}$ results from a combination of atmospheric density scaling from envelope fractions and mean-molecular weight effects as well as being shaped by mass-loss. For a mixed envelope, a more steep slope is expected for lower metallicity envelopes and vice-versa.
	
	\item Finally, the radius valley characteristics are further important quantities to reproduce. We introduced our approach to quantifying this aspect in Sects. \ref{sec:st_mass_dependency} and \ref{sec:radius_valley_sol}.
	
\end{itemize}
\begin{figure*}
	\centering
	\includegraphics[width=.88\linewidth, trim=0 7 0 7., clip]{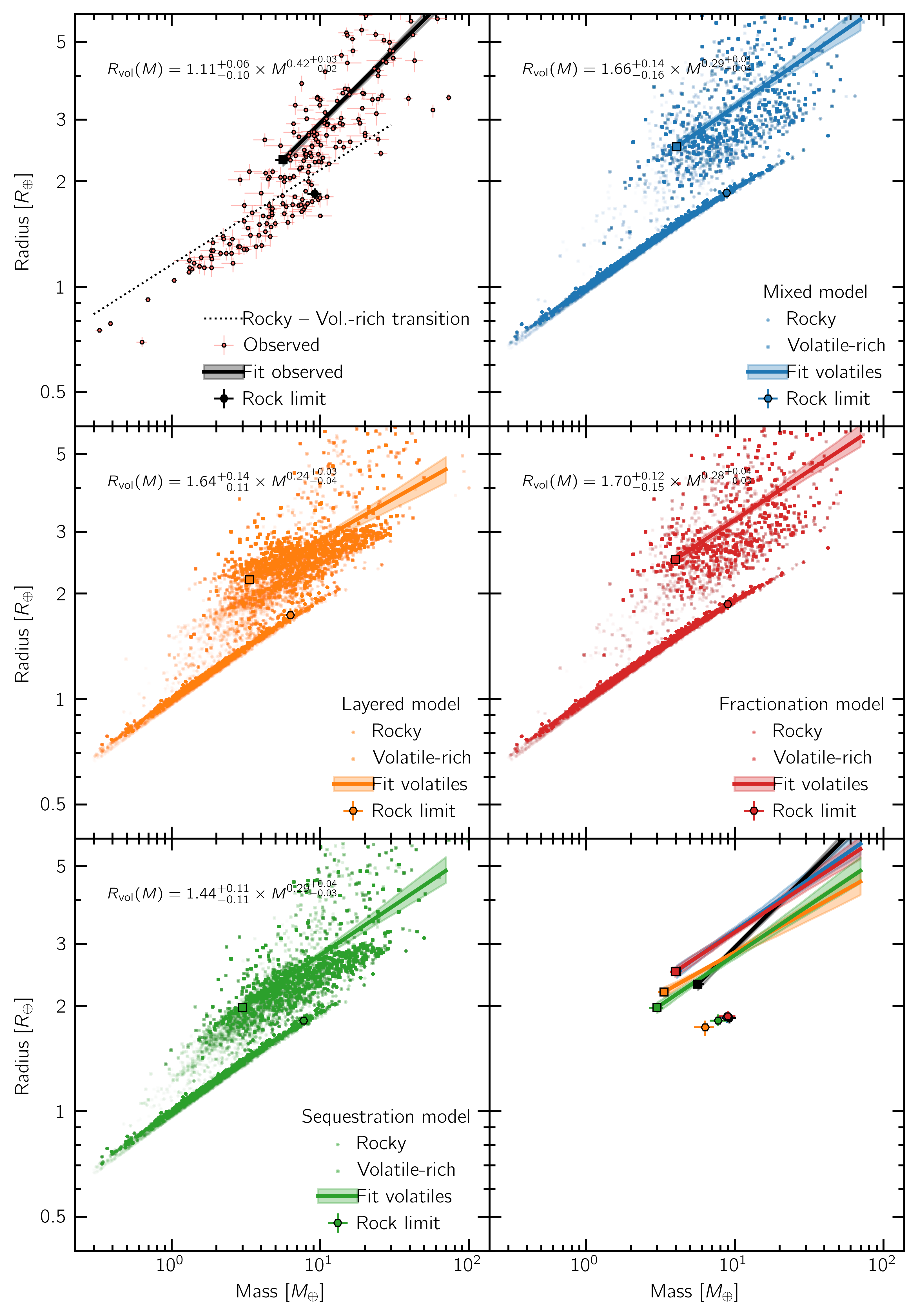}
	\caption{Comparison of observed exoplanets and different model results in the mass radius plane. The synthetic data includes the estimate of the observational bias as detailed in Sect. \ref{sec:sampling}. A power-law was fitted to the volatile-rich populations for each panel and indicated in the top left using units of Earth radii and masses (see text for details). The lower end is drawn at the 10th quantile of the volatile distribution ($M_{\rm vol, min}$). Furthermore, the 90th quantile of the rocky planet population ($M_{\rm rock, max}$) is indicated with errors obtained from the same resampling technique. The bottom right panel shows repeated fits for the different populations for better comparison to each other.}
	\label{fig:m_r_models_fits}
\end{figure*}

Fig. \ref{fig:m_r_models_fits} provides a means for comparison of mass-radius distributions with corresponding power-law fits to the volatile distribution as well as visualizing the quantities $M_{\rm rock,max}$ and $M_{\rm vol, min}$. We find that all models predict a smaller value than the observed $M_{\rm vol, min}$, match within errors $M_{\rm rock,max}$, and predict a significantly shallower slope of the volatile planet mass-radius relation $m_{\rm vol,\,M-R}$ than the observed $m_{\rm vol,\,M-R}$. We find that the quantity $M_{\rm vol, min}$ is in our model shaped by mass-loss and not imprinted from formation: The population of water-rich planets after formation extends to lower masses than $M_{\rm vol, min}$ (i.e. to approximately 1\,M$_\oplus$). Since modeled $M_{\rm vol, min}$ lie below the observed one, this hints at too inefficient photoevaporation caused by the prescription itself or by too high mean molecular weight atmospheres in the modeled planets.

While it can be a useful probe for photoevaporation, the values for $M_{\rm vol, min}$ could be influenced by the observational bias. In contrast, the slope of the mass-radius relation is a less bias-dependent indication that too many planets with high metallicity envelopes are present in all models. This can also be seen from a bifurcation in the volatile-rich population with a group clustered at relatively low radii and another with larger radii. The lower population is made from almost pure steam worlds. Although, some planets populate this region (e.g. GJ 9827 d, \citealp{Passegger2024}), they are not as commonly observed as predicted by all models. This group is more abundant in the Water Sequestration and Layered models compared to the Mixed one.

The qualitative observation that $M_{\rm vol, min} < M_{\rm rock, max}$ is captured by all models. If all rocky planets originate from being stripped from an equal-composition massive envelope, there should be an irradiation trend visible with rocky planets close to $M_{\rm rock, max}$ orbiting on short orbital periods and, vice-versa, volatile-rich planets around $M_{\rm vol, min}$ on wide orbits. We visually inspect the data for such a trend and recover it for the Mixed model but do not find any correlation in the observed data.

\begin{figure*}
	\centering
	\qquad\qquad\large{\textsf{$M_\star$ below 0.6\,M$_\odot$}}\qquad\qquad\qquad\qquad\qquad\large{\textsf{$M_\star$ larger than 0.6\,M$_\odot$}}
	\includegraphics[width=.85\linewidth, trim=0 7 0 7., clip]{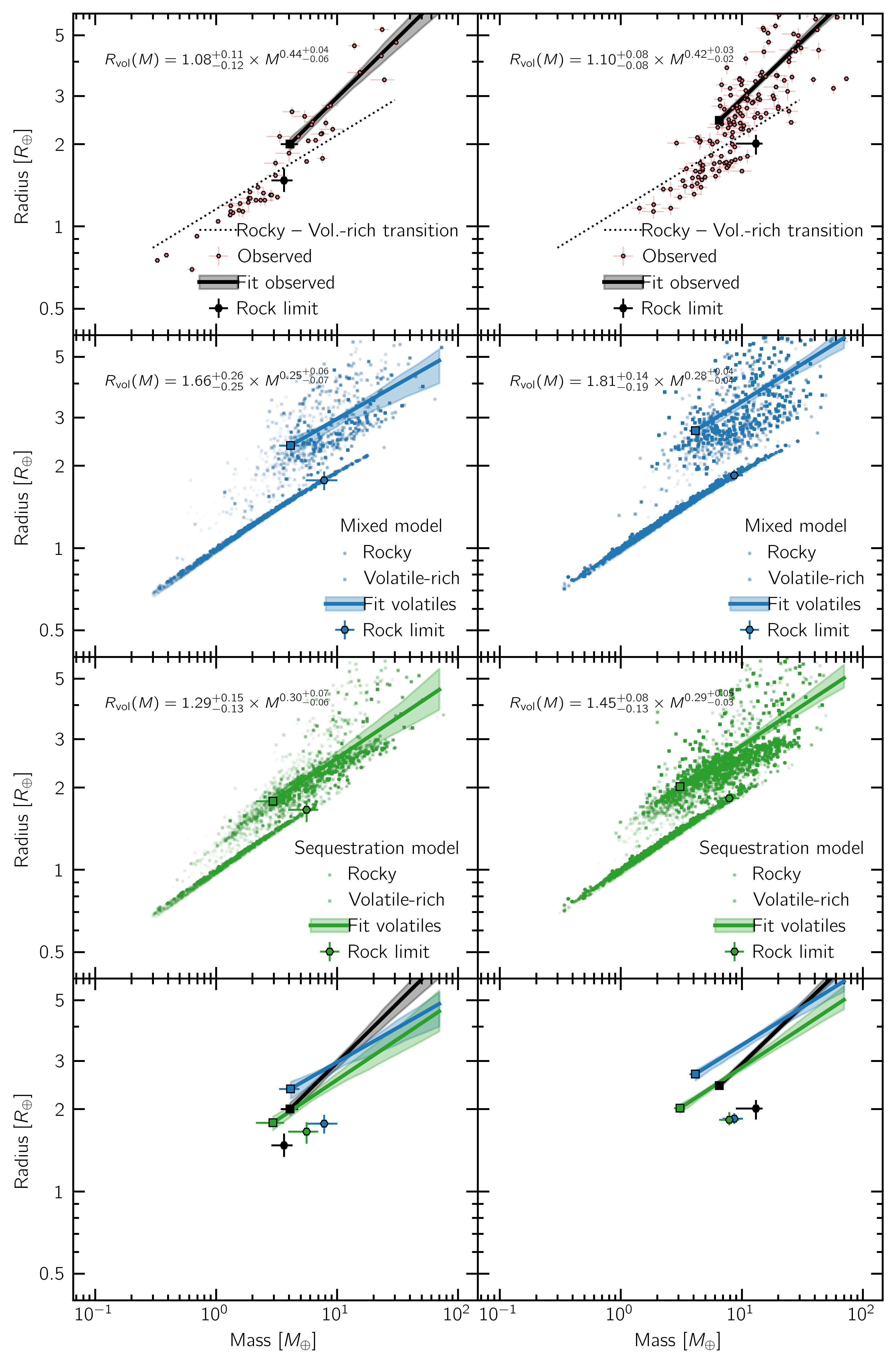}
	\caption{Comparison of mass-radius relations with observational data split by stellar mass. Data is presented and fitted equivalently to Fig. \ref{fig:m_r_models_fits}. The left column shows simulations and observations with $M_{\star}<0.6\,$M_${\odot}$ while the panels on the right show the data for more massive stars.}
	\label{fig:m_r_fits_st_mass}
\end{figure*}

We also examine mass–radius trends with stellar mass by splitting the observational data and the Mixed and Water Sequestration model output into two bins, separated at 0.6\,M$_\odot$ (Fig. \ref{fig:m_r_fits_st_mass}). One apparent effect is that the two characteristic masses $M_{\rm vol, min}$ and $M_{\rm rock,max}$ lie closer to each other around lower stellar masses for the synthetic and observed data. Instead of an overlap in mass space of the volatile-rich and rocky population, they show a clear transition at $\sim$4\,M$_\oplus$. In the Mixed model, this is mainly because the rocky planet population decreases in mass. It is to a large degree sourced by planets which formed interior to the iceline and this population shrinks with decreasing stellar mass \citep{Burn2021}. The findings here seem to contradict the theoretical findings of the work by \citet{Venturini2024}. They found for their pebble-accretion based planets (with structures analogous to the Mixed model) that the overlap of the volatile-rich planets decreases with increasing stellar mass -- opposite to what we find. However, this difference is because we applied a cut at the runaway greenhouse limit to our synthetic planets (Sect. \ref{sec:sampling}). The planets with radii larger than that expected from condensed water are also not present around low mass stars in the work of \citet[][their Fig. 1]{Venturini2024}. Indeed, when excluding planets cold enough to allow for condensed water, their data reveals a similar transition from rocky to volatile-rich planets at 3-4\,M$_\oplus$ around low-mass stars.

The comparison suggests that our Mixed model could be improved if planets have lower initial envelope metallicities: the slope $m_{\rm vol,\,M-R}$ would steepen and the $M_{\rm vol, min}$ would increase towards a generally better match with observations. This is partially the case for the highest masses in the Sequestration model (see Sect. \ref{sec:results_seq}). Other possible options are that less water or more H/He should be accreted during formation. Within the standard assumptions on planet formation, the latter option is unlikely as the formation model already includes optimistic estimates on the envelope cooling efficiency (via reduced assumed opacities, \citealp{Mordasini2014}) which should not lie lower and thus cannot increase the H/He content at fixed mass. However, it is possible to accrete also more H/He at fixed mass if part of the metals reside in the envelope (fully consistent with the Mixed model, see \citealp{Venturini2016} and the discussion in Sect. \ref{sec:formation_mixing}).

Potential mechanisms for a reduction of water accretion could be radioactive heating and subsequent loss of water from planetesimals \citep{Lichtenberg2019} or recycling of water back to the ambient gas \citep{Moldenhauer2022,Wang2023}. However, with a reduction of the atmospheric metallicity the radius valley location would shift to even larger radii \citep{Mordasini2020} in disagreement with Kepler data (see Sects. \ref{sec:st_mass_dependency} and \ref{sec:radius_valley_sol}). Thus, the process which reduces the water content might require a planet mass-dependency which does not affect low-mass planets but more higher mass ones, such as expected for recycling \citep{Wang2023}.

\subsection{Negligible effect of fractionation}
\label{sec:fractionation_interpretation}
In this work, we included the effect of fractionating outflows of oxygen relative to hydrogen for the evolution of synthetic exoplanets (see Appendix \ref{app:fractionation} for the implementation). This effect was studied in the context of exoplanets and coupled to evolutionary calculations by \citet{Hu2015}, \citet{Malsky2020,Malsky2023}, \citet{Gu2023}, and \citet{Cherubim2024} although for He or deuterium. Since oxygen is heavier, the effect should be more pronounced. However, we did not find a significant effect on the radius.

Since the initial envelope masses are considerably massive, following the output of our formation modeling, a large photoevaporative mass loss is required to alter their radius significantly. Such a flow is however not in the regime where fractionation occurs as can be seen from Eq. \eqref{eq:fractionation}, where the mass loss $\Phi_{\rm H}$ is in the denominator of the subtrahend. Thus, for $\Phi_{\rm H} \gg (m_{\rm O} - m_{\rm H}) b_{\rm H,O} \frac{G M}{R^2 k_{\rm B} T_{\rm iso}}$, which evaluates to a relative mass loss of $\dot{M_{\rm H}}/M \gg \SI{5e-13}{1/yr}$, fractionation does not occur $x_{\rm O} \approx 1$ and mass of oxygen is lost proportionally to its fraction \citep{Hunten1987,Zahnle1986,Zahnle1990}. While this estimate is based on analytical theory, the numerical investigation of \citet{Johnstone2020} found a similar estimate of this so-called critical flux. This is satisfied for most planets in our simulation until the last stage of the evolution (see Sect. \ref{sec:evo_track}) in general agreement with the findings of \citet{Cherubim2024}. However, in that study, He-rich atmospheres remained in some cases. Although this difference should be investigated in more details, it could be due to their study also sampling small initial envelope fractions on the percent level. This speculation is supported by the study of \citet{Gu2023} who found a dependency on the initial gas envelope mass for deuterium fractionation. Such low envelope fractions are not obtained from our formation modeling and are therefore not included in this study but could be candidates for fractionation to occur to a more significant degree.

We do not find a significant influence of the estimate of the upper atmosphere number density (see Appendix \ref{app:fractionation}). Further extension of the fractionation model to different species, such as He, and including a realistic temperature and flow profile should be investigated in the future \citep[see also the discussions of deviations from analytical estimates in][]{Johnstone2020,Schulik2023}.

There are observational prospects to probe the atmospheres of planets at the upper edge of the radius valley \citep{Malsky2023,Cherubim2024}. Since He and deuterium are unlikely to be outgassed from the interior, they provide a probe for the process of fractionation occurring. However, a future observational confirmation of the absence of He-dominated atmospheres should not be over-interpreted as being caused by the complete absence of photoevaporation. Instead, the planet could have had a more massive initial gaseous envelope and photoevaporation could be efficiently acting and also remove the He content of the planet. Rather, the presence of He or deuterium rich atmospheres could be used as a probe for initial gaseous envelope masses.

\subsection{Outlook on models including sequestration}
\label{sec:sequestration_discussion}
Here, we report comparisons to observational data which do not favor the most physically advanced Water Sequestration model, at least at planetary masses below 3\,M$_\oplus$. However, this finding should be put into relation to the developments required for a comprehensive model with atmosphere-interior coupling.

First, the Water Sequestration model includes only the dissolution of water but excludes that of the other modeled constituents, namely He and hydrogen. For He, this could be justified, but hydrogen dissolution into the interior is expected to occur \citep{Hirschmann2012,Kite2019}. Moreover, the dissolved material is bound to undergo chemical reactions. Instead of outgassing water, the secondary envelope can be of a different composition \citep[see the recent works by][for equilibrium chemistry results]{Werlen2025a,Werlen2026a}. However, if large quantities of oxygen were accreted via water ice and, correspondingly, the mantle is in a high redox state, water is indeed expected to also be the main constituent of the outgassed atmosphere \citep{Bower2022,Lichtenberg2023}. Nevertheless, given how sensitive photoevaporative mass-loss is on the radius and therefore the molecular weight of the atmosphere, more outgassed species should be included. In particular, gas accretion predicts that the hydrogen to oxygen ratio increases with increasing planetary mass \citep{Mordasini2016}, therefore, the evolution of the high-mass planets in our simulation would change more significantly and outgassed hydrogen and other species could improve the match with observations in this regime.

Another neglected effect is the temperature dependency of the solubility. For simplicity, we assumed a constant exterior temperature of 3000\,K for the computation of the water sequestration. Instead, the planet would cool and the retention of water in the magma ocean would therefore decrease \citep{Chachan2018}. This would imply earlier outgassing than found here and potentially more efficient loss of water from the planet. Of a similar order of magnitude but with inverse effects is the simplification that pressures used for sequestration include only the water envelope mass and ignore the H/He content. A consistent pressure at the magma surface would slightly increase water fractions in the melt for planets with significant H/He content. We note that the presence of water on planets resulting from the formation model is usually either negligible or significant in mass which then leads to water being the dominant constituent in the atmosphere, and thus determining the total pressure.

A potentially more important related effect is however the solidification of the magma ocean \citep{Hamano2013,Bower2018,Bower2022}. In our approach, the whole mantle material is assumed to remain in molten state. However, depending on the details of crystallization pattern and viscosity (e.g. whether or not a solid lid forms at the top of the mantle), most of the volatile budget could be removed from the mantle upon solidification. If it occurred at an early stage on hot planets up to three Earth masses, this effect might in fact reconcile with observations the low-mass end of the Water Sequestration model results presented here since outgassing would decrease the bulk density of the planet leading to more efficient, likely complete, loss of volatiles and thus removal of planets from the low-density, low-mass region where no real planets are found. On the other hand, outgassing of volatiles increases opacities and prevents cooling, thereby counteracting the solidification of the magma. Therefore, our assumption of a continued existence of magma during the early stage might be justified. In contrast, if solidification and increased outgassing occurs at a late stage when H/He was lost and photoevaporation rates are reduced, outgassing could decrease the bulk density and worsen the mismatch of the simulated planets' densities to observations.

The model assumes fully connected reservoirs. However, this assumption may overestimate the efficiency of volatile transport to the deep interior. Within the magma ocean, mixing of water from the atmosphere to the interior could potentially be inhibited if the lower density of water-rich melts causes a buoyant, compositionallly stratified layer (\citealp{Solomatova2021}, see also \citealp{Spaargaren2020}). Furthermore, convection in the solid mantles of super-Earths could be less efficient than in Earth-sized planets due to adiabatic compression or differences in rheology at high pressures \citep{Karato2011,Miyagoshi2017,Miyagoshi2018,Shahnas2018}. Combined, these effects could limit the transport of volatile elements into the deep interior, emphasizing the need to better resolve planetary interior phases, viscosity, and thermal state.

An exploration in this direction is also motivated by the findings of \citet{Gaidos2024} who explore the age dependency of the small planet population. They suggest that the atmosphere-interior coupling could produce effects on Gyr timescales as observed and further motivate an inclusion of water condensation. Qualitatively, water condensation could have a similar effect as switching from the Mixed to the Layered scenarios and could be quantified using these approaches as a basis \citep[see also the discussion in][]{Venturini2024}.

\subsection{Mixed-composition atmosphere photoevaporation}
\label{sec:discussion_photoevaporation}
Our approach to photoevaporation of an atmosphere containing mixtures of gases follows a simple mass-weighted approach (Eq. \ref{eq:Mdot_total}). This approach would only hold if gases would evaporate independently of each other. However, as emphasized by the discussion in Sect. \ref{sec:fractionation_interpretation}, gases are well coupled in the regime of significant mass loss. Furthermore, even trace amounts of some gases with strong absorption in the infrared wavelength range, such as water, can strongly affect the temperature structure in the atmosphere and thus the mass-loss rate \citep{Yoshida2022}. As such, the photoevaporative mass loss of mixtures cannot be treated independently and a pure mass weighted approach is too simplistic to fully represent the exact photoevaporation rate of mixed gases. In particular, hydrogen escape is more efficient by almost an order of magnitude even when comparing a planet of equal mass and radius leading to a significant increase in mass loss at $Z\sim$0.9 (a relatively typical value in our simulations) compared to a pure H$_2$O based on the prescription used here. We note that we chose this approach due to the limited availability of alternatives in the literature at the start of the project.

However, recently, \citet{Yoshida2025a} calculated escape rates as a function of water to hydrogen ratio. Their results are comparable to those of our employed rates from \citet{Johnstone2020} in the high-water content limit. In the low-H$_2$O/H$_2$ regime, the rates differ significantly from those tabulated by \citet{Kubyshkina2021}, although mostly in the low-gravity regime not explicitly studied and fitted by \citet{Yoshida2025a}.\footnote{The fitting function provided by \citet{Yoshida2025a} and compared here assumes a functional form motivated by energy-limited escape. However, \citet{Kubyshkina2021} show that the scaling in the low-gravity regime deviates from energy-limited escape.}

\begin{figure}
	\includegraphics[width=\linewidth]{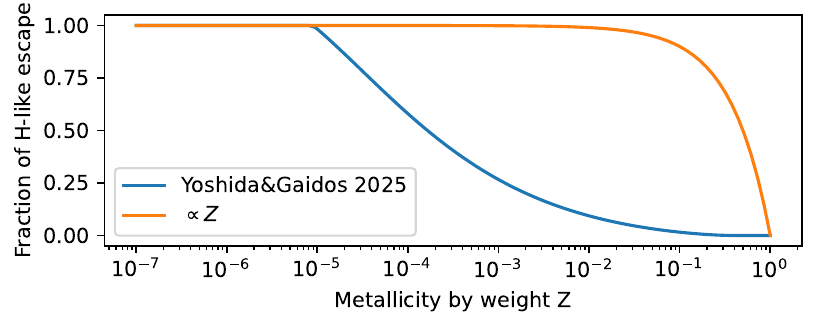}
	\includegraphics[width=\linewidth]{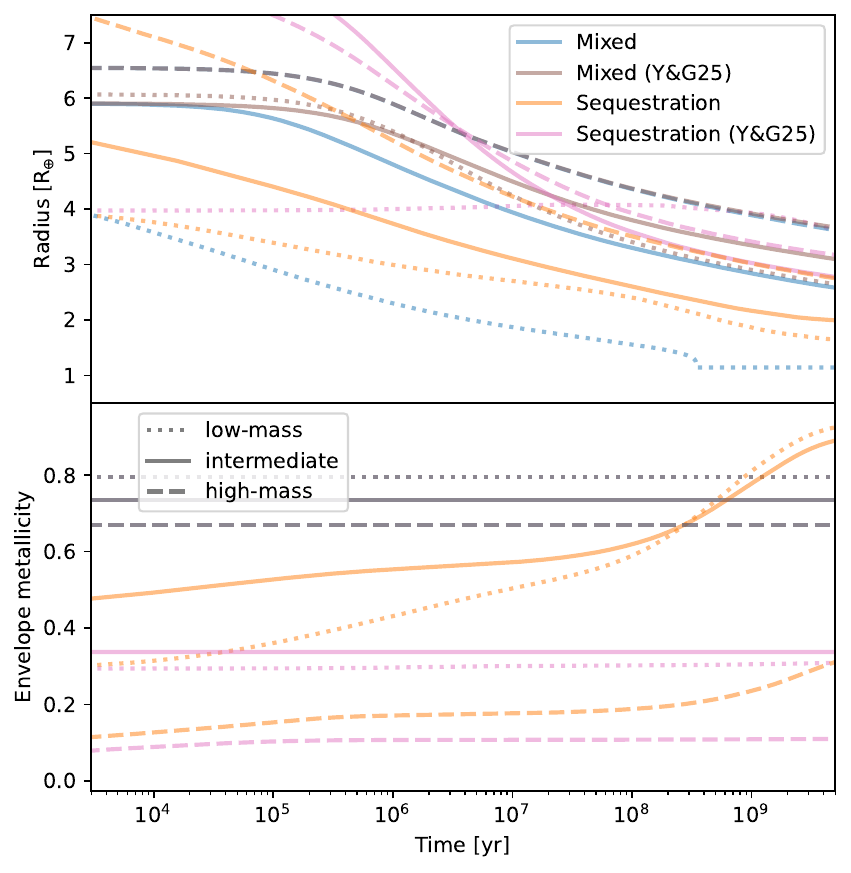}
	\caption{Comparison of metallicity scaling (top) and radius and metallicity evolution tracks (bottom) using the photoevaporative escape rates found by \citet{Yoshida2025a} and the ones used in this work. The vertical axis in the top panel corresponds to the different combinations used here (Eq. \ref{eq:Mdot_total}) compared to \citet{Yoshida2025a} of H/He-like mass loss rates with those of pure water (1 indicating $\dot{M}_{\rm evap} = \dot{M}_{\rm H/He}$, 0 $\dot{M}_{\rm evap} = \dot{M}_{\rm H_2O}$). As Fig. \ref{fig:evo_tracks}, the lower figure contains a radius and envelope metallicity evolution panel and nominal Mixed and Sequestration cases.}
	\label{fig:YoshidaGaidosComparison}
\end{figure}

For mixed compositions, driven by strong cooling of H$_2$O in the infrared regime, \citet{Yoshida2025a} find a strong decrease of mass-loss at low contents. As cooling saturates due to the atmosphere becoming locally optically thick in the relevant wavelength bands, a saturation towards high H$_2$O/H$_2$ follows. Combined, these results indicate a scaling different from a linear dependency on $Z_{\rm env}$ (Fig. \ref{fig:YoshidaGaidosComparison}, top)\footnote{For simplicity, we treated He as H}.

For a comparison of the effect which retains the low-gravity regime by \citet{Kubyshkina2021}, we re-scaled the fitting function given by \citet{Yoshida2025a} to the unit interval [0,1] for the dependency on H$_2$O/H$_2$ (by number of molecules) and used it in place of the linear dependency on $Z_{\rm env}$ in Eq. \eqref{eq:Mdot_total}, such that H$_2$O/H$_2=0$ and 1 recover the \citet{Kubyshkina2021} and \citet{Johnstone2020} cases, respectively. With this approach, we aim to test the sensitivity of the model on the scaling with composition while leaving a full adoption of the escape rates found by \citet{Yoshida2025a} to future works.

The lower panel of Fig. \ref{fig:YoshidaGaidosComparison} shows a significant dependency on the used approach. With a scaling which more strongly weighs the lower H$_2$O escape rates, and essentially converges to the pure water case at $Z_{\rm env}=0.1$, the water-rich cases investigated here ($Z_{\rm env}$ ranging from 0.1 to 0.9) are subject to less photoevaporative mass-loss. In particular, the low-mass case is affected and retains its atmosphere. Also affected by the lower mass-loss is the coupling to the sequestration cases, which reside at lower $Z_{\rm env}$ initially and will evolve much less in overall mass and thus in water content of the core and atmosphere.

The implication on the mass-radius distribution is based on these estimates expected to be significant and has mixed effects on the match to observations. At low mass, planets would be more likely to retain atmospheres and worsen the match to observations with the mass-radius distribution of water-rich sub-Neptunes extending to lower masses. For more massive planets, retention of mixed-composition envelopes would be favored with a revised scaling. Therefore, the mass of the most massive stripped bare core ($M_{\rm rock,max}$ discussed in Sect. \ref{sec:fits}) would decrease and approach the most massive dry super Earth formed (as in the Layered and Sequestration cases). This, in turn, would agree better with observations. For the sequestration model, the radii at low masses would increase due to less loss of H/He. Furthermore, we expect an effect on the slope of the mass-radius relation of sub-Neptunes. However, it is without extended calculations difficult to speculate whether the slope should increase due to a vanishing effect of $Z_{\rm env}(M_{\rm planet})$ or decrease due to lower mass-loss rates for all simulated sub-Neptunes residing in the efficiently H$_2$O cooled regime ($Z_{\rm env}>0.1$). For both the Mixed and Sequestration models, lower escape rates also move the radius valley to lower radii and more planets would retain atmospheres increasing the number of sub-Neptunes compared to super Earths. This latter effect might improve the match to observations for the Mixed model (see Sect. \ref{sec:radius_valley_sol}).

The large amplitude of the deviation between escape models and their strong impact on the population of exoplanets motivates further investigations and the development of comprehensive grids of escape rates for inclusion in global models as a function of composition in general -- as it is not clear that water is the right constituent (Sect. \ref{sec:sequestration_discussion}, as well as e.g. \citealp{Werlen2026a}). For this purpose, we suggest the use of mantle redox state or atmospheric C/O as dimension in escape rate studies, which would require the use of an (equilibrium) chemistry model coupled to the hydrodynamics and radiation transport.

\subsection{Other Uncertainties and Caveats}
\subsubsection{Formation model parameters}
\label{sec:discussion_formation_parameters}
When synthesizing planetary populations, several parameters are not tightly constrained. Therefore, different pathways can lead to populations with reasonable reproduction of the observed exoplanet bulk mass, total radius, and distance distributions. A review highlighting this challenge, listing parameter choices in the literature, and highlighting their effect on the planet population was recently compiled \citep{Burn2024a}. In particular, the migration prescription combined with the chosen disk initial conditions and viscosity sensitively affects whether the observable small planet population contains migrated planets or those formed in-situ.

The formation simulations extended here assume a viscous disk with standard prescriptions for type I migration from \citet{Paardekooper2011} using the same viscosity as for the disk evolution. The discussion on whether and to which degree disks can be modeled as viscous (i.e. their mechanisms driving turbulence can be modeled with an effective viscosity) is ongoing. There is a realistic possibility that inner disk regions ($\sim 1-50\,$au) turn relatively laminar during extended stages in disk evolution. This would likely trigger earlier gap-opening of planets with the effect of reduced migration of sub-Neptune mass planets (e.g. \citealp{Ogihara2018}, see also discussion in \citealp{Burn2024a}). We note that the creation of migration prescriptions for low-viscosity disks is an active research topic and need to be included in future global simulations for more accurate conclusions \citep{Paardekooper2023,Ziampras2024,Aoyama2023,Weder2025}.

While more efficient migration, or similarly, a more shallow radial distribution of planetesimals, promotes the abundance of water-rich sub-Neptune planets, it does not have a direct impact on the pattern in mass-radius space as the growth history of the individual planets remains the same. In the analysis done here, with the exception of Fig. \ref{fig:biasedcumulativeobservedmixedlayeredfractionationburn-2024a}, we did not put strong emphasis on the number of sub-Neptunes compared to super Earths.

The changes in mass radius space, in our simulations, are governed by the planetary composition and the partitioning of elements discussed here. It is therefore important to review the effects on bulk composition of a planet, which is naturally related to the initial composition of the disk. For example, the maximum water abundance in solids is limited by the disk water abundance -- which in turn is, in the outer disk, governed by the partitioning of the available oxygen to icy species. The observed ice abundances towards star forming regions show significant scatter \citep{Boogert2015}. Here, they were not varied between stars (the initial abundances follow \citealp{Marboeuf2014b}) although there could be significant intrinsic scatter. Furthermore, the stellar abundance was kept Solar and the oxygen fraction in minerals was similarly fixed \citep[following][]{Thiabaud2014}. Moreover, we assumed that no refractory organics are present. The addition of some refractory carbon species (pure amorphous carbon or organic compounds such as PAH) is strongly favored by the observed carbon depletion in diffuse interstellar clouds \citep{Whittet2003}. Depending on the composition and ratio of carbon to oxygen in the refractory organics, the ice abundance can vary. If carbon is condensed as amorphous carbon, water abundance might be promoted, but more likely is a depletion of all ices (including water) due to condensation into refractory organics. Furthermore, the water abundance could be negatively affected by physical processes (radioactive heating, recycling) discussed above.

Finally, the results presented here are in principle sensitive to the bulk abundance of H/He to water on the planet. This ratio is sensitive to the efficiency of gas accretion which in turn depends on energy input (i.e. solid accretion rate), boundary conditions varying with the location in and model of the disk, and assumed envelope opacities. We note that here, we chose low envelope opacities during formation, calibrated using a grain growth model by \citet{Mordasini2014}. However, as discussed in \citet{Mordasini2020}, \citet{Venturini2024}, and other locations in this manuscript, there is a balancing effect in evolutionary calculations with photoevaporation: for mixed envelopes and planets with significant water content, larger H/He contents reduce the envelope metallicity and inflate the planet size. This leads to more efficient mass loss of the water-rich envelopes and finally smaller final planet sizes compared to a case with equal water and lower H/He content. Due to these competing effects, at short orbital periods, a higher H/He content leads to smaller final planets while at larger distances (or larger planetary mass), the opposite effects holds. Another effect of more efficient H/He accretion would be that planets more easily reach runaway gas accretion and the formation of giants would be promoted, which in turn requires a re-tuning of other parameters in a population synthesis exercise to match the observed abundance of giants.

Overall, we expect the mass-radius relationship -- in shape -- to have relatively small sensitivity on typical planet formation parameters which keep the composition fixed but be rather sensitive to gas accretion efficiency and assumptions on solid compositions (e.g. ice to dust ratio). However, the exact abundances of planets in the mass radius distribution (e.g. number of sub-Neptunes to super Earths) is sensitive to parameters affecting migration rates and growth efficiencies such as viscosity and properties of the solid planet building blocks. Future works should disseminate these dependencies in more detail.

\subsubsection{Mixing in the formation stage}
\label{sec:formation_mixing}
In this work, we assume for the Mixed model that mixing occurs after the formation stage. For ease of implementation and computational efficiency, we did not mix H/He with water during the assembly of the planets in the formation stage. If mixing occurs, it likely emerges as soon as certain threshold conditions are met \citep{Innes2023}. A disk-embedded planet with a mixed envelope structure has different properties compared to a layered structure. Its higher density implies that the mass resides deeper in the potential well which favors gas accretion while cooling could be negatively affected by higher molecular opacities due to water vapor. If opacities remain dust-dominated and are therefore not sensitive to water opacity, a positive effect on gas accretion results \citep{Venturini2015,Venturini2016,Kimura2020}. The same effect is expected if the structure is fully convective, as is likely for enriched planets with enhanced atmospheric densities \citep{Ikoma2001,Hori2011}. This implies that the H/He to water ratio in the planets modeled here could be too low and future model generations need to allow for consistent treatment of the partitioning during formation and evolution. This could help to improve the match with observations for the Mixed model as discussed in Sect. \ref{sec:discussion}. 

\subsubsection{Collisions}
Giant impacts among young planets are expected to significantly affect the planets. Depending on impact mass ratios and velocities, different outcomes can be expected, such as disruption of the bodies, merging, or no efficient mass transfer (hit-and-run) \citep[e.g.][]{Cambioni2019}. In terms of bulk composition this leads to compositional mixing with more intermediate-composition planets emerging. This aspect is included in the simulations presented here where perfect merging of all components except H/He envelopes is assumed. However, atmospheres present are also significantly affected \citep{Schlichting2015,Schlichting2018}. For this aspect, we could not use a fully consistent treatment of collisions even though we use multiple-body simulations as basis (see Sect. \ref{sec:structure_and_compo}). This has two impacts on our results.

First, atmospheric mass-loss is likely underestimated in cases where collisions occurred after the disk dissipation. While the atmosphere of the smaller collision partner is always removed from the mass budget, we assumed no effect on the mass of the atmosphere of the target. \citet{Venturini2020} discuss in detail the effect of using a model for atmospheric mass removal due to collisions and found a general shift of the population towards higher densities.

Second, for the Layered model, mixing of the atmosphere is likely to occur after a giant impact. This would imply that the planet transitions from a Layered scenario to the Mixed case. From our simulations, we expect this effect to affect the population differently depending on their atmospheric escape rates. If the timescale of mass-loss is shorter than the few Myr on which the collisions occur, H/He is already lost at the time at which the impacts occur. This is the case for the close-in, low-mass population which is completely stripped of H/He and is making up most of the observed population below $\sim 5\,$M$_\oplus$. For the planets where atmospheres are Mixed by impacts, they would become more resistant to H/He loss. It is noteworthy that giant collisions are also common during the disk stage which further promotes the Mixed scenario over the Layered one.

\section{Summary and conclusions}
\label{sec:conclusion}
We investigated the masses and radii of planets resulting from coupled formation and evolution modeling under the assumption that water can be abundant on sub-Neptune and super Earth mass exoplanets. The underlying formation model relies on planetesimal accretion and includes both initially rocky and water-rich planets in the observable regime. The large water fraction on planets with masses ranging from Earth mass to tens of Earth masses is a prediction of planet formation models including orbital migration and was in a prior work found to match reasonably well the radius distribution of observed exoplanets around Solar-type stars \citep{Burn2024}. Here, we followed-up on this work and explore the mass-radius relations of these planets. Furthermore, we slightly revised the hydrogen and helium content after formation and extend the simulations to lower stellar masses and discussed trends with stellar mass. In \citet{Burn2024}, the water was mixed with H/He in the gaseous envelope or alternatively segregated from H/He and layered on top of the rocky core. Here, we additionally explore the effect of preferential removal of hydrogen over oxygen (fractionation) and sequestration of water into the interior of the planet while improving the physical treatment of the layered scenario (i.e. by using a realistic equation of state). 

The following conclusions can be drawn:
\begin{itemize}
	\item We found a qualitatively good match of the observational data if water is mixed with H/He in the envelope and is not allowed to dissolve in the interior. In contrast, planets with layered structures do not match observed mass-radius relations.
	\item For the planets with mixed envelope composition, the location of the radius valley is increasing for stellar masses ranging from 0.1 to \SI{0.5}{\msol} before remaining constant. The observed slope with orbital period is reproduced within errors. The locus of the radius valley is found to be at slightly too large radii and the sub-Neptune population is not numerous enough, which did not occur for the initial conditions with less H/He used in \citet{Burn2024}. However, we recover the trend of longer minimal orbital periods for the sub-Neptunes compared to super Earths, which was not found in \citet{Burn2024}, thanks to enforcing the presence of H/He on all planets at the end of the disk lifetime. Another iteration of modeling is required to match both aspects.
	\item The obtained mass-radius relation is in general agreement with the coupled formation-evolution modeling by \citet{Venturini2024}. The differences are due to different orbital period distributions and applied limits as well as due to their pebble-accretion-informed initial conditions segregating more distinctly water-rich from rocky planets compared to the prediction of planetesimal accretion used here. As in that work, the lowest-mass volatile-rich planets are predicted at lower irradiation levels, while a cut at fixed irradiation predicts a stellar-mass independent emergence of the lowest-mass sub-Neptune.
	\item The quantification of mass-radius characteristics suggests that at high planetary mass $\sim 20\,M_{\oplus}$, a mechanism should act during formation which moderately lowers the initial water content compared to standard formation modeling. This could be a reduction of the overall budget -- via atmospheric recycling with the gaseous disk or $^{26}$Al heating and water-loss on planetesimals -- or by sequestration of water to the interior for which there are hints at a better match to observations at high planetary masses.
	\item For smaller planets shaping the Kepler radius valley, assuming the photoevaporation model is accurate, metallicities intermediate between the low values of fully mixed primordial envelopes and the high values of a layered envelope or one in equilibrium with a magma ocean are required.
	\item Related to the too small locus of the radius valley, the Water Sequestration model predicts a population of low-mass ($M<3\,M_{\oplus}$) planets with water abundances close to unity in the gaseous envelope. These water-worlds do not seem to be present in significant numbers in nature \citep{Parviainen2024}. This might indicate that less than maximum sequestration or earlier outgassing occurs. The model used here is likely too simplistic for a conclusive assessment and we suggest to include time-dependent solidification of the assumed persistent magma ocean and comprehensive tracking of other chemical species and elements as well as their reactions \citep{Werlen2025a,Werlen2026a} in future works. This is especially motivated by a promising match to observations at higher planetary masses.
	\item Mass loss driven by photoevaporation on observable exoplanets is found to be mostly equally removing hydrogen and oxygen for the modeled planets because all planets with surviving volatile budgets initially contain large envelopes which require large mass loss rates. This implies, that fractionation is not necessarily occurring on sub-Neptunes and the absence of enriched (e.g. He-rich) planets cannot generally be used as a probe to answer whether evolution or formation determines the present-day exoplanet properties. Instead, occurrence or absence of fractionation in nature constrains the initial volatile budget of the planet. We note that recent, more detailed fractionation models provide additional nuance to this conclusion \citep{Valatsou2026}.
	\item The linear combination of photoevaporative mass-loss models of pure compositions might overestimate actual mass loss rates and thus underestimate final radii of mixed composition envelopes and should be more extensively tabulated and included in future generation of evolution models.
\end{itemize}

The conclusions here rely on a heterogeneous sample of exoplanetary masses and radii. In addition to the discussed model improvements, we strongly encourage the construction of a homogeneous sample of well characterized planets or occurrence rates in mass-radius-orbital period space, which is, for example, being pursued by the THIRSTEE project \citep{Lacedelli2024} and which should also be possible from the future PLATO mission and follow-up.

\paragraph{Data availability}
Tables \ref{tab:data_mixed}-\ref{tab:data_seq} are available on Zenodo \url{https://doi.org/10.5281/zenodo.22960286}.

\begin{acknowledgements}
	We thank J. Venturini, C. Mordasini, Th. Henning, H. Klahr, and E. Pall\'e for fruitful discussion. We further thank the contributions and discussions of all participants of the "Density Matters" Ringberg workshop which helped to shape this work. Finally, we thank the anonymous referee for insightful comments which helped to improve the quality of analysis and manuscript.
	
	R.B. acknowledges the financial support from DFG under Germany’s Excellence Strategy EXC 2181/1-390900948, Exploratory project EP 8.4 (the Heidelberg STRUCTURES Excellence
	Cluster) and from the European Research Council (ERC) under the European Union’s Horizon 2020 research and innovation program under grant agreement No. 832428-Origins and via the ERC Advanced Grant “TiPPi –
	Turbulence, Pebbles and Planetesimals: The Origin of Minor Bodies in the Solar
	System” (project ID 855130, PI: Klahr).
	
    R.L. is funded by the European Union (ERC, THIRSTEE, 101164189) and acknowledges financial support from the Severo Ochoa grant CEX2021-001131-S funded by MCIN/AEI/10.13039/501100011033 and by NASA through the NASA Hubble Fellowship grant \#HST-HF2-51559.001-A awarded by the Space Telescope Science Institute, which is operated by the Association of Universities for Research in Astronomy, Inc., for NASA, under contract NAS5-26555. 
    
    C.D. acknowledges support from the Swiss National Science Foundation under grant TMSGI2_211313.
	
	Parts of this work has been carried out within the framework of the NCCR PlanetS supported by the Swiss National Science Foundation under grants 51NF40\_182901 and 51NF40\_205606.
	
	Calculation were performed on the \textit{Horus} cluster at the University of Bern and on the computation facilities at the Max Planck Institute for Astronomy.
	
	The plots shown in this work were generated using \texttt{matplotlib} \citep{Hunter2007} and the analysis made use of the \texttt{scikit-learn} \citep{scikit-learn} and \textit{scipy} \citep{scipy} packages.
\end{acknowledgements}

\bibliographystyle{aa_url}
\bibliography{library_betterbib_zotero} 

\begin{appendix}
	\section{An analytic model for fractionation}
	\label{app:fractionation}
	For large escape rates, the analytic estimate by \citet{Hunten1987} should apply \citep{Zahnle1990}. They derived a fractionation factor
	\begin{equation}
	x_{\rm O} = 1 - \frac{\left(m_{\rm O} - m_{\rm H}\right) b_{\rm H,O} G M}{\Phi_{\rm H} R^2 k_{\rm B} T_{\rm iso} \left(1+ \frac{n_{\rm O}(R)}{n_{\rm H}(R)}\right)}\,,
	\label{eq:fractionation}
	\end{equation}
	where $M$ is the mass of the planet, $T_{\rm iso}$ is the temperature in the atmosphere assumed in this model to be isothermal, $b_{\rm H,O} = \SI{4.8e17}{\per\centi\meter\per\second} \left( T_{\rm iso}/\SI{1}{\kelvin}\right)^{0.75}$ is the binary diffusion coefficient \citep{Mason1970,Zahnle1986,Pavlov2019}, $R$ is the effective planetary radius, and $\Phi_{\rm H}$ (respectively $\Phi_{\rm O}$) is the escaping flux of hydrogen (oxygen) particles per surface such that the total particle loss $\dot{N}_{\rm H} = - 4\pi R^2 \Phi_{\rm H}$ ($\dot{N}_{\rm O} = - 4\pi R^2 \Phi_{\rm O}$).
	
	The oxygen fractionation factor is defined to relate to the fluxes as
	\begin{equation}
	x_{\rm O} \equiv \frac{\Phi_{\rm O}}{\Phi_{\rm H}} \frac{N_{\rm H}}{N_{\rm O}}\,,
	\end{equation}
	where the total number of particles of both species can be obtained from the total masses used in the rest of the evolution model $N_{\rm O} = M_{\rm O}/m_{\rm O}$ and $N_{\rm H} = M_{\rm H}/m_{\rm H}$. The number density ratio $n_{\rm O}/n_{\rm H}$ at radius $R$ is more challenging to obtain as discussed below. This factor is relevant for major abundances of the heavier oxygen and whenever the mass loss rate $\Phi_{\rm H}$ is relatively low.
	
	After some algebra and relating these to the total mass loss $\dot{M}_{\rm evap} = \dot{M}_{\rm H} +  \dot{M}_{\rm O}$, the expression can be written as
	\begin{equation}
	\dot{M}_{\rm O} = \frac{M_{\rm O}}{M} \left( \dot{M}_{\rm evap} - \frac{4 \pi m_{\rm H} (m_{\rm O} - m_{\rm H}) b_{\rm H,O} G M}{k_{\rm B} T_{\rm iso}  \left(1+ \frac{n_{\rm O}(R)}{n_{\rm H}(R)}\right)} \right)
	\end{equation}
	implying (given that the amount of water is determined by the available oxygen and never limited by hydrogen availability) a water loss of
	\begin{equation}
	\dot{M}_{\rm H_{2}O} = \dot{M}_{\rm O} \frac{m_{\rm H_{2}O}}{m_{\rm O}}\,.
	\end{equation}
	
	To investigate the effect of different estimates of $n_{\rm O}/n_{\rm H}$, we assume uniform mixing $n_{\rm O}(R)/n_{\rm H}(R) = N_{\rm O}/N_{\rm H}$ and $T_{\rm iso} = \SI{3000}{\kelvin}$ in one case (lower limit for fractionation) and an estimate assuming isothermal atmospheres and no mixing as detailed in the following to give a more optimistic estimate for fractionation.
	
	For the optimistic fractionation model, we determine the base of the flow analogous to literature on disk photoevaporation \citep{Hollenbach1994} by requiring that the mass flux launched as a decoupled wind at the absorption layer of high-energy radiation \citep{Murray-Clay2009} is sustained from a deeper 'base' radius $R_{\rm base}$ of the flux. This can be written as
	\begin{equation}
	\dot{M}_{\rm evap} = c_{\rm s, ions} 4 \pi R^2_{\rm base} \rho_{\rm base}\,,
	\end{equation}
	where  $c_{\rm s, ions} = \frac{k_{\rm B} \SI{3000}{\kelvin}}{\bar{m}}$, with $\bar{m} = m_{\rm H} \left(X/0.5 + Y/2 + Z/3 \right)^{-1}$ for mass fractions of hydrogen $X$, helium $Y$ and heavy elements assumed to be water $Z$. This averaging assumes that particles are singly ionized. For example H$_2$O is present as $2\times$H$^+$, O$^+$, and 3 electrons with a total weight of 18\,u distributed to 6 particles, thus an average weight of 3\,u per particle results. The sound speed is an order-of-magnitude estimate of average velocity with which the ionized particles in the wind are launched. The base radius can be located outside of the resolved structure of the envelope which sometimes requires extrapolation. Then, the number density ratio $\frac{n_{\rm O}(R_{\rm base})}{n_{\rm H}(R_{\rm base})}$ is estimated assuming individual hydrostatic distributions for O and H with scale heights $H_{\rm O, H} = \frac{k_{\rm B} T}{m_{\rm O, H}}$. This implies no interaction between O and H, thus a lower limit for $\frac{n_{\rm O}}{n_{\rm H}}$, respectively an upper limit for the fractionation effect. More detailed considerations should include different species \citep{Odert2018}, the effect of turbulent mixing \citep[e.g.][]{Charnay2015,Komacek2019}, ionization and the resulting interactions with ions \citep{Hu2015,Guo2019}, and the deviations of detailed numerical studies from the analytical approach \citep{Johnstone2020,Schulik2023}.
	
	\section{Iron core and mantle densities including volatiles}
	\label{app:core_density_effect_water_in_core}
	The Water Sequestration model presented here simplifies the treatment of the interior structure to determine the radius of the iron core and magma ocean by assuming a water layer on top of the rocky material instead of consistently mixing water with mantle and core. In this section, we aim at comparing the radii of the planets obtained in this way with a more consistent model based on \citet{Dorn2017}. 
	
	This model focuses on Earth-like rocky interiors with steam atmospheres. Water can exist in the core, mantle, or surface, depending on the thermal state of the planet. The core is composed of iron (Fe) with light elements such as hydrogen (H) and oxygen (O). For solid Fe, we use the equations of state for hexagonal close-packed iron from \citet{Hakim2018}, and for liquid iron and its alloys, we follow \citet{Luo2024} who use a Mie-Gr\"uneisen equation of state with parameters derived from ab-initio calculations. \citet{Luo2024} state that their approach is valid in the super-Earth to sub-Neptune regime, that is for high pressures above 50\,GPa. The core is assumed to be adiabatic, but there is a temperature jump at the core-mantle boundary (CMB) due to residual heat from core formation, based on \citet{Stixrude2014}.
	
	The mantle consists of three primary components: MgO, SiO$_2$, and FeO, assuming an Earth-like composition. For solid mantle properties, we employ the Perple\_X thermodynamic model \citet{Connolly2009} and the database of \citet{Stixrude2022}. For pressures above $\sim$125 GPa, stable minerals are defined a priori using equations of state from various sources \citep{Fischer2011, Faik2018, Hemley1992, Musella2019,Coppari2021}. The liquid mantle’s density is computed as an ideal mixture of Mg$_2$SiO$_4$, SiO$_2$, and FeO, using Mg$_2$SiO$_4$ instead of MgO due to the recent data for forsterite being updated for high-pressure regimes.
	
	The mantle is considered fully adiabatic, with water present only in the melts, while solid mantle is assumed to be dry. Water decreases the mantle's density following \citet{Bajgain2015}, and this reduction is nearly pressure and temperature independent for small water fractions. The mantle melting curve is calculated for pure MgSiO$_3$, and the addition of water \citep{Katz2003} and iron \citep{Dorn2018} lowers the melting temperature. Water in the core also reduces the melting temperature, as modeled by \citet{Luo2024}. The water which is not dissolved in the mantle (see Sect. \ref{sec:structure_and_compo}) is modeled as being in steam or supercritical phases using the AQUA equation of state \citep{Haldemann2020}, with an isothermal profile below 0.1 bar that transitions to an adiabatic profile.
	
	Hydrated silicates are not considered motivated by a minor impact on the density \citep{Shah2021}. Instead, water is treated using the EOS of \citet{Haldemann2020} and the additive volume law to combine it with rock. In this way, interior structures and planetary core and mantle radii can be obtained comparable to the simplified model adopted in this work.
		
	\begin{figure}[h]
		\centering
		\includegraphics[width=\linewidth]{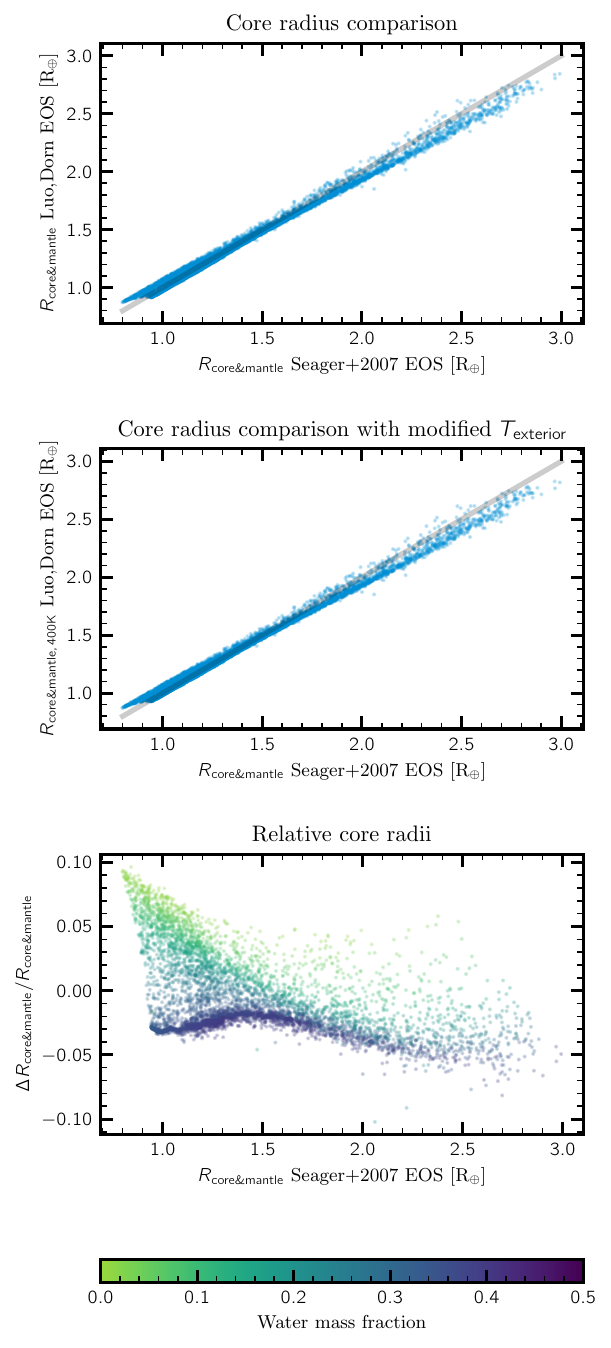}
		\caption{Comparison of radii resulting from two different interior structure calculations in the Water Sequestration scenario without outgassing. The adopted interior structure model uses the equation of states (EOS) based on \citet{Seager2007} (see Sect. \ref{sec:structure_and_compo}) and assumes internally that water forms a layer of ice on top of the rocky material. The resulting radii of the solid planetary core and mantle are shown on the horizontal axis. Those from a model based on \citet{Luo2024,Dorn2021}, which calculates more consistently the density of materials including water (see text), is shown on the vertical axis (top and middle panel). It assumes an exterior temperature of 3000\,K (top panel), respectively 400\,K (middle). The bottom panel shows the relative difference between the two modeled radii, with positive values indicating larger radii and total water mass fraction color-coded.}
		\label{fig:radii_comparison}
	\end{figure}

	The results of this comparison for the planets which contain some water but without H/He envelopes from our simulations can be seen in Fig. \ref{fig:radii_comparison}. We see a difference of up to 10\% at low mass and low water fraction mainly influenced by the differing assumed rocky mantle material. For substantial water fractions, the radii obtained with the consistent interior model are lower by up to 5\% compared to the same planets with a layered structure. We find a compensation of the effect of the varying rocky material density and the treatment of water. Isolated, the assumption that water is layered on top of the mantle instead of mixed leads to a decrease in radius of 15\%. We conclude that for our main purpose, the evolution of the planetary atmosphere, this difference is likely not substantial to change the outcome significantly also thanks to the compensation effects. A change in core radius on the order of 5\% (15\%) leads to a 15\% (50\%) change in atmospheric mass loss for planets whose radius is dominated by the core. However, for detailed comparison to observations and retrieval of planetary parameters, our radii should not be used.

	\section{Supporting figures}
	\label{app:supp_figures}
	This appendix contains figures used for more detailed and higher-dimensional analysis of the synthetic populations presented in the main part of this work.
	
	\begin{figure*}
		\centering
		\includegraphics[width=\linewidth]{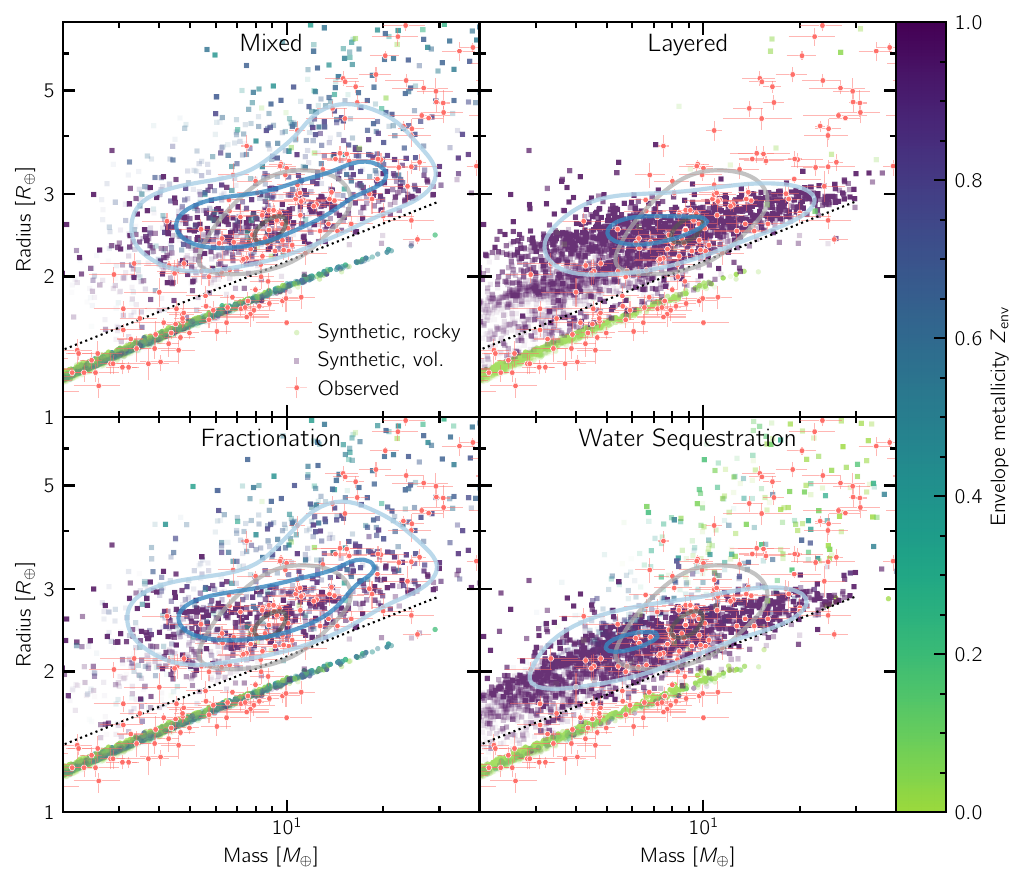}
		\caption{Zoom into sub-Neptune region of the mass radius relation. Similar to Fig. \ref{fig:m_R_all}, but here the envelope metallicity $Z_{\rm env}$ is coded in color after 5\,Gyr of evolution. The coloring of rocky planets corresponds to the metallicity of the last gaseous envelope before removal. In addition, the plot contains the contours of a two-dimensional kernel density estimate \citep[KDE][]{scikit-learn} of the probability density of volatile rich planets (bulk densities below 0.65 the expected Earth density, \citealp[][]{Zeng2016}) for both observed (gray) and synthetic (blue). To account for uneven observation errors, the KDE uses a bandwidth of 0.1 measured in logarithms base 10 and $e$ for masses and radii, respectively.}
		\label{fig:m_R_zhomo_kde}
	\end{figure*}

	\begin{figure}
		\centering
		{\large \textsf{Layered}}
		\includegraphics[width=\linewidth]{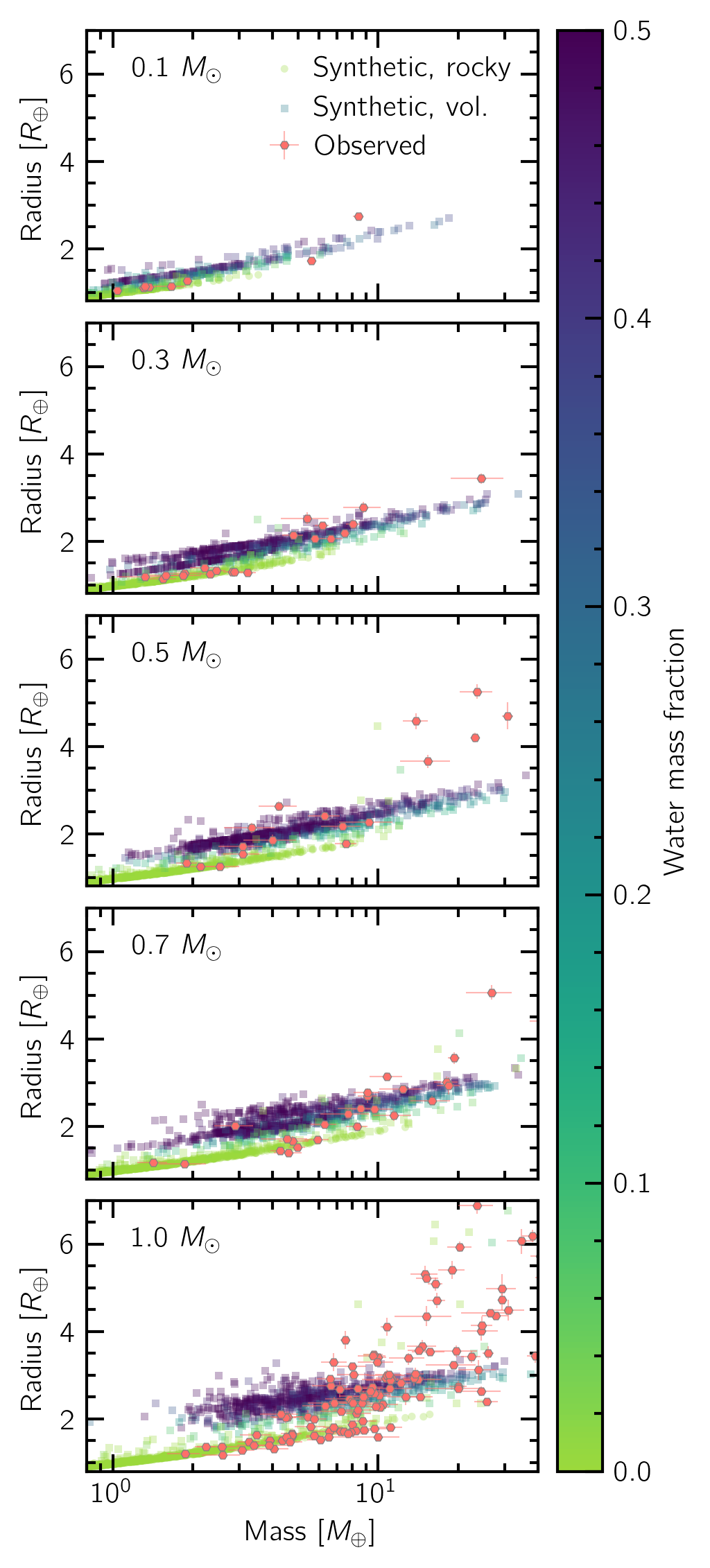}
		\caption{As Fig. \ref{fig:m_r_stmasses_mixed} but showing the output of the Layered model. The bifurcation of water-rich (blue colored) planet densities at low masses, best seen in the 0.3\,M$_\odot$ panel, originates from the phase transition of water. The panels here include for the 0.1 and 0.3\,M$_\odot$ cases planets exterior to the runaway greenhouse limit where water condenses. With increasing stellar mass, as temperatures increases, so does the scatter in water envelope radii at fixed planetary mass.}
		\label{fig:m_r_stmasses_layered}
	\end{figure}

	\begin{figure}
	\centering
		{\large \textsf{Water Sequestration}}
		\includegraphics[width=\linewidth]{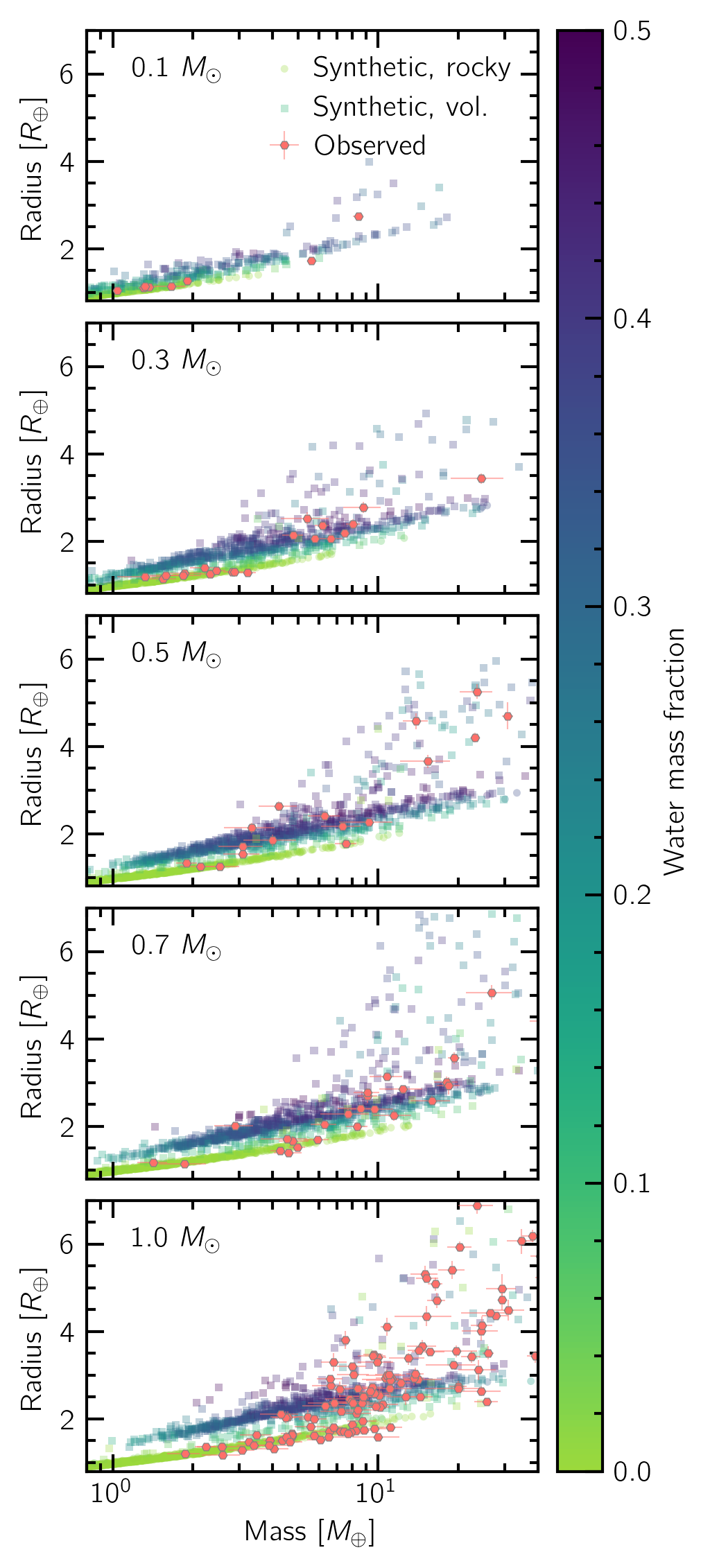}
		\caption{As Fig. \ref{fig:m_r_stmasses_mixed} but showing the output of the Water Sequestration model.}
	\label{fig:m_r_stmasses_waterincore}
	\end{figure}

	\begin{figure}
		\centering
		\includegraphics[width=\linewidth, trim=0 33 0 7., clip]{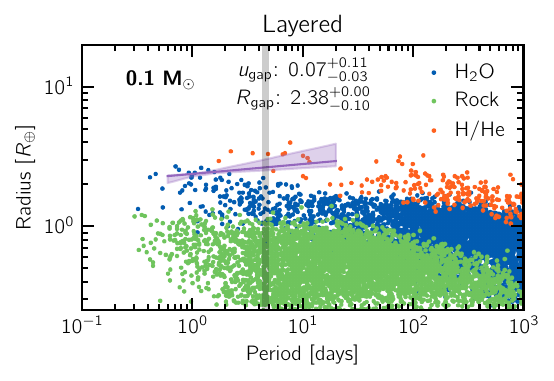}
		\includegraphics[width=\linewidth, trim=0 33 0 7., clip]{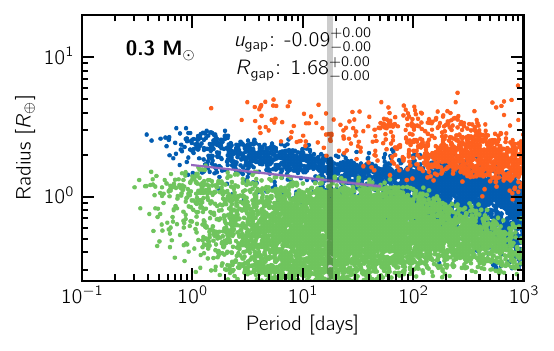}
		\includegraphics[width=\linewidth, trim=0 33 0 7., clip]{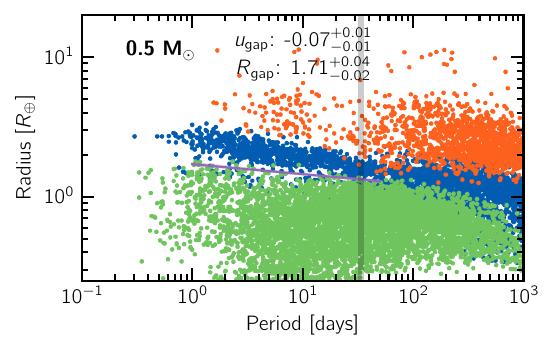}
		\includegraphics[width=\linewidth, trim=0 33 0 7., clip]{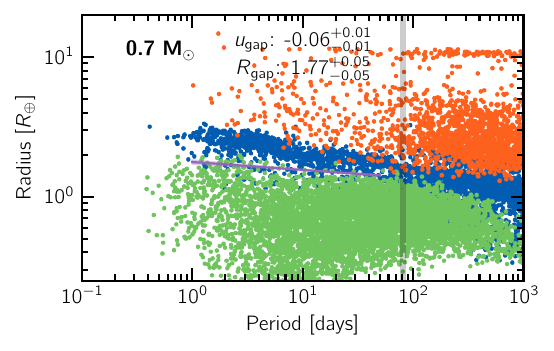}
		\includegraphics[width=\linewidth, trim=0 7 0 7., clip]{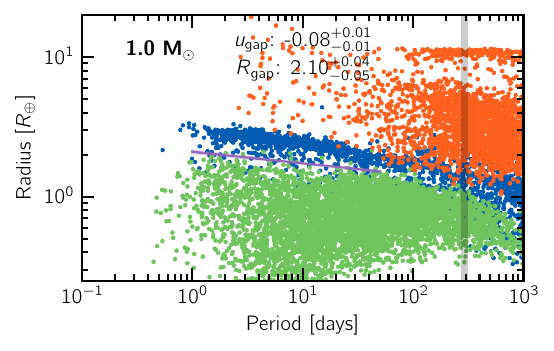}
		\caption{Period versus radius of unbiased, synthetic planets using the layered model around stars of different masses. The gap was fitted analogous to the Mixed model results shown in Fig. \ref{fig:period_radius_Mstars}. We suggest to not use the 0.1\,M$_\odot$ case due to low-number statistics.}
		\label{fig:period_radius_Mstars_layered}
	\end{figure}

	\begin{figure}
		\centering
		\includegraphics[width=\linewidth, trim=0 33 0 7, clip]{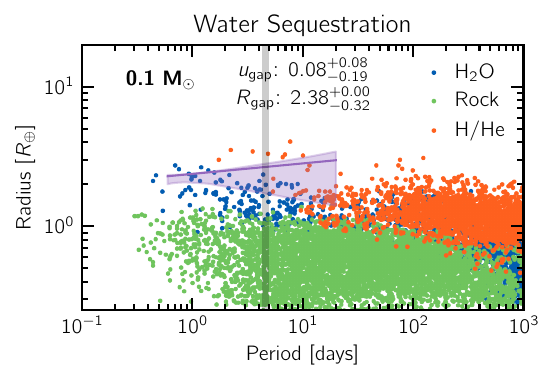}
		\includegraphics[width=\linewidth, trim=0 33 0 7., clip]{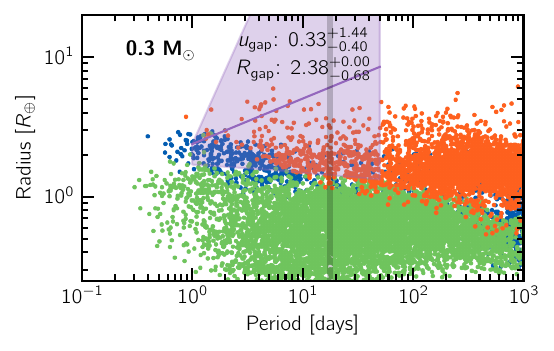}
		\includegraphics[width=\linewidth, trim=0 33 0 7., clip]{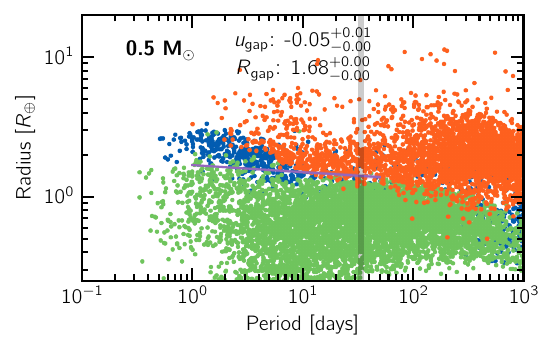}
		\includegraphics[width=\linewidth, trim=0 33 0 7., clip]{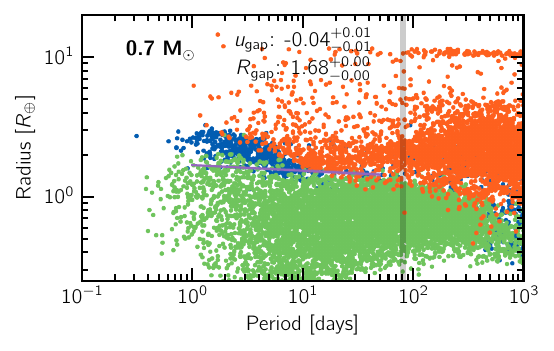}
		\includegraphics[width=\linewidth, trim=0 7 0 7., clip]{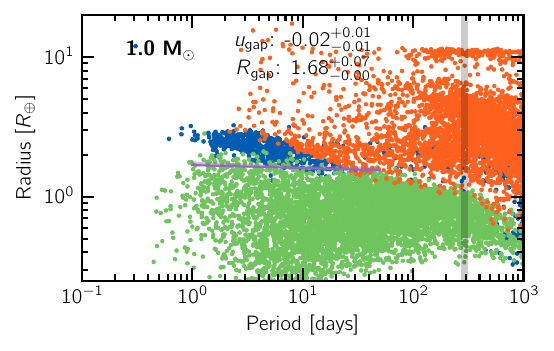}
		\caption{Period versus radius of unbiased, synthetic planets using the Water Sequestration model around stars of different masses. The gap was fitted analogous to the Mixed model results shown in Fig. \ref{fig:period_radius_Mstars}. We caution the use of the 0.1\,M$_\odot$ and 0.3\,M$_\odot$ fits.}
		\label{fig:period_radius_Mstars_water_in_core}
	\end{figure}

\section{Supporting Tables}
Here, we provide tables of simulated planets in the regime where our assumptions apply, which is considered to be within the runaway greenhouse limit (see Sect. \ref{sec:sampling}). The tables provided here do not use an estimate of the observational bias and therefore include every simulated planet once irrespective of detection likelihood. The full tables are available in electronic form \href{https://doi.org/10.5281/zenodo.22960286}{online}.

\newpage
\begin{landscape}
\begin{table}
	\caption{Simulated planet parameters for the Mixed model}
	\label{tab:data_mixed}
	\begin{tabular}{r | c | c | c | c | c | c | c | c | c | c | c | c | c | c | c | c  }
		\hline\hline
		ID & $M$ [M$_{\oplus}$] & $R$ [R$_{\oplus}$] & $a$ [au] & $P$ [days] & $F$ [F$_{\oplus}$] & $i$ & $e$ & $M_{\rm env}$ [M$_\oplus$] &$M_{\rm core}$ [M$_\oplus$] & $Z_{\rm env}$ & $f_{\rm H_{2}O}$ & sysID & pltID & $M_\star$ [M$_\odot$] & $T_{\rm eff}$ [K] & $P_{\rm rgh}$ [days]\\
		\hline
		0 & 0.9870 & 0.9744 & 0.0177 & 2.7170 & 0.2200 & 5.24e-7 & 0.028 & 0.0000 & 0.9870 & 0.772 & 0.0000 & 4 & 11 & 0.1 & 2810 & 4.600 \\
		1 & 1.7341 & 1.3627 & 0.0061 & 0.5461 & 1.8685 & 5.24e-7 & 0.019 & 0.0987 & 1.6354 & 1.000 & 0.0569 & 4 & 21 & 0.1 & 2810 & 4.600 \\
		2 & 1.3143 & 1.0589 & 0.0215 & 3.6345 & 0.1493 & 5.24e-7 & 0.021 & 0.0000 & 1.3143 & 0.698 & 0.0000 & 4 & 38 & 0.1 & 2810 & 4.600 \\
		\hline
	\end{tabular}
	\tablefoot{The full table with column descriptions is available in electronic form.}
\end{table}

\begin{table}
	\caption{Simulated planet parameters for the Layered model}
	\label{tab:data_layered}
	\begin{tabular}{r | c | c | c | c | c | c | c | c | c | c | c | c | c | c | c | c  }
		\hline\hline
		ID & $M$ [M$_{\oplus}$] & $R$ [R$_{\oplus}$] & $a$ [au] & $P$ [days] & $F$ [F$_{\oplus}$] & $i$ & $e$ & $M_{\rm env}$ [M$_\oplus$] &$M_{\rm core}$ [M$_\oplus$] & $Z_{\rm env}$ & $f_{\rm H_{2}O}$ & sysID & pltID & $M_\star$ [M$_\odot$] & $T_{\rm eff}$ [K] & $P_{\rm rgh}$ [days]\\ \hline
		0 & 1.3249 & 1.4414 & 0.0177 & 2.7170 & 0.2200 & 5.24e-7 & 0.028 & 0.3379 & 0.9870 & 1.000 & 0.2550 & 4 & 11 & 0.1 & 2810 & 4.600 \\
		1 & 1.7425 & 1.3695 & 0.0061 & 0.5460 & 1.8690 & 5.24e-7 & 0.019 & 0.1072 & 1.6354 & 1.000 & 0.0000 & 4 & 21 & 0.1 & 2810 & 4.600 \\
		2 & 2.1690 & 1.4828 & 0.0215 & 3.6344 & 0.1493 & 5.24e-7 & 0.021 & 0.8547 & 1.3143 & 1.000 & 0.3940 & 4 & 38 & 0.1 & 2810 & 4.600 \\	
		\hline
	\end{tabular}
	\tablefoot{The full table with column descriptions is available in electronic form.}
\end{table}

\begin{table}
	\caption{Simulated planet parameters for the Fractionation model}
	\label{tab:data_frac}
	\begin{tabular}{r | c | c | c | c | c | c | c | c | c | c | c | c | c | c | c | c  }
		\hline\hline
		ID & $M$ [M$_{\oplus}$] & $R$ [R$_{\oplus}$] & $a$ [au] & $P$ [days] & $F$ [F$_{\oplus}$] & $i$ & $e$ & $M_{\rm env}$ [M$_\oplus$] &$M_{\rm core}$ [M$_\oplus$] & $Z_{\rm env}$ & $f_{\rm H_{2}O}$ & sysID & pltID & $M_\star$ [M$_\odot$] & $T_{\rm eff}$ [K] & $P_{\rm rgh}$ [days]\\ \hline
		0 & 0.9870 & 0.9742 & 0.0177 & 2.7170 & 0.2200 & 5.24e-7 & 0.028 & 0.0000 & 0.9870 & 0.772 & 0.0000 & 4 & 11 & 0.1 & 2810 & 4.600 \\
		1 & 1.7423 & 1.3705 & 0.0061 & 0.5460 & 1.8692 & 5.24e-7 & 0.019 & 0.1070 & 1.6354 & 1.000 & 0.0614 & 4 & 21 & 0.1 & 2810 & 4.600 \\
		2 & 1.3143 & 1.0589 & 0.0215 & 3.6345 & 0.1493 & 5.24e-7 & 0.021 & 0.0000 & 1.3143 & 0.698 & 0.0000 & 4 & 38 & 0.1 & 2810 & 4.600 \\
		\hline
	\end{tabular}
	\tablefoot{The full table with column descriptions is available in electronic form.}
\end{table}

\begin{table}
	\caption{Simulated planet parameters for the Sequestration model}
	\label{tab:data_seq}
	\begin{tabular}{r | c | c | c | c | c | c | c | c | c | c | c | c | c | c | c | c  }
		\hline\hline
		ID & $M$ [M$_{\oplus}$] & $R$ [R$_{\oplus}$] & $a$ [au] & $P$ [days] & $F$ [F$_{\oplus}$] & $i$ & $e$ & $M_{\rm env}$ [M$_\oplus$] &$M_{\rm core}$ [M$_\oplus$] & $Z_{\rm env}$ & $f_{\rm H_{2}O}$ & sysID & pltID & $M_\star$ [M$_\odot$] & $T_{\rm eff}$ [K] & $P_{\rm rgh}$ [days]\\ \hline
		0 & 1.1104 & 1.1793 & 0.0177 & 2.7170 & 0.2200 & 5.24e-7 & 0.028 & 0.0045 & 1.1059 & 1.000 & 0.1110 & 4 & 11 & 0.1 & 2810 & 4.600 \\
		1 & 1.7374 & 1.2851 & 0.0061 & 0.5463 & 1.8677 & 5.24e-7 & 0.019 & 0.0016 & 1.7358 & 1.000 & 0.0587 & 4 & 21 & 0.1 & 2810 & 4.600 \\
		2 & 1.6181 & 1.3616 & 0.0215 & 3.6345 & 0.1493 & 5.24e-7 & 0.021 & 0.0100 & 1.6081 & 1.000 & 0.1876 & 4 & 38 & 0.1 & 2810 & 4.600 \\
		\hline
	\end{tabular}
	\tablefoot{The full table with column descriptions is available in electronic form.}
\end{table}
\end{landscape}
	
\end{appendix}

\end{document}